\documentclass{aastex6}

\newcommand\devolx{(1+z)$^{1.6}$}

\newcommand\levolx{(1+z)$^{1.5}$}
\newcommand\epeak{E$_{\rm peak}$}
\newcommand\epi{E$_{\rm pi}$}

\newcommand\eiso{E$_{\rm iso}$}
\newcommand\liso{L$_{\rm iso}$}
\newcommand\zmax{z$_{\rm max}$}
\newcommand\Pf{P$_{\rm f}$}
\newcommand\dur{T$_{90}$}

\newcommand\kngrb{69}
\newcommand\fngrb{52}
\newcommand\orngrb{95}
\newcommand\fkngrb{26}
\newcommand\fncomp{34}
\newcommand\kncomp{59}
\newcommand\fmedz{1.85}
\newcommand\kmedz{1.77}

\begin{document}


\title{The Maximum Isotropic Energy of Gamma-Ray Bursts}

%% Use \author, \affil, plus the \and command to format author and affiliation 
%% information.  If done correctly the peer review system will be able to
%% automatically put the author and affiliation information from the manuscript
%% and save the corresponding author the trouble of entering it by hand.
%%
%% The \affil should be used to document primary affiliations and the
%% \altaffil should be used for secondary affiliations, titles, or email.

%% Authors with the same affiliation can be grouped in a single
\author{J-L.~Atteia\altaffilmark{1}, V.~Heussaff, J.-P.~Dezalay, A.~Klotz and D.~Turpin\altaffilmark{2}}
\affil{Universit\'e de Toulouse; UPS-OMP; IRAP; Toulouse, France\\ 
CNRS; IRAP; 14, avenue Edouard Belin, F-31400 Toulouse, France}

\author{A.E.~Tsvetkova and D.D.~Frederiks}
\affil{Ioffe Institute, Politekhnicheskaya 26, St. Petersburg, 194021, Russia}

\author{Y.~Zolnierowski}
\affil{LAPP, Universit\'e de Savoie, CNRS/IN2P3, 9 chemin de Bellevue, BP 110, 74941 Annecy-le-Vieux, France}

%\and

\author{F.~Daigne and R.~Mochkovitch}
\affil{UPMC-CNRS, UMR7095, Institut d'Astrophysique de Paris, F-75014, Paris, France}

%\author{Julie Steffen\altaffilmark{4}}
%\affil{UPMC-CNRS, UMR7095, Institut d'Astrophysique de Paris, F-75014, Paris, France}
%
%%% Use the \and command so offset the last author.
%\and
%
%\author{Jeff Lewandowski\altaffilmark{5}}
%\affil{IOP Publishing, Washington, DC 20005}


\altaffiltext{1}{jean-luc.atteia@irap.omp.eu}
\altaffiltext{2}{D. Turpin used funds provided by the LabEx OCEVU.}
%\altaffiltext{3}{AAS Journals Associate Editor-in-Chief}
%\altaffiltext{4}{AAS Director of Publishing}
%\altaffiltext{5}{IOP Senior Publisher for the AAS Journals}

\begin{abstract}

The most energetic gamma-ray bursts (GRBs) are remarkable sources releasing huge amounts of energy on short timescales.
Their prompt emission, which usually lasts few seconds, is so bright that it is visible across the whole observable universe.
Studying these extreme events may provide clues on the nature of GRB progenitors and on the physical processes at work in relativistic jets.
%They are also crucial to understand the evolution of GRBs, in density or luminosity, since they are visible out to large redshifts. 

In this paper, we study the bright end of the isotropic energy distribution of long GRBs. 
We use two samples of long GRBs with redshift detected by \textit{Fermi}/GBM or Konus\textit{--Wind}, two instruments which measure the spectral shape and the energetics of the prompt emission accurately. 
We focus on GRBs within a range of redshifts z = 1 -- 5, a volume that contains a large number of energetic GRBs, and we propose a simple method to reconstruct the bright end of the GRB energy distribution from the observed one.
We find that the GRB energy distribution cannot be described by a simple power law but requires a strong cutoff above $1-3 \times10^{54}$ erg. 
We attribute this feature to an intrinsic limit on the energy per unit of solid angle radiated by gamma-ray bursts.

\end{abstract}




\section{Introduction}
\label{sec_intro}
Gamma-Ray Bursts are extremely energetic sources, which can release isotropic energies (\eiso \footnote{The isotropic energy is computed under the assumption that the source emits isotropically.}) in excess of {$10^{54}$ erg} in gamma-rays.
We investigate here the bright end of the GRB energy distribution with the purpose of determining whether it contains indications of a limit to the energy that GRBs radiate in gamma-rays.

The GRB energy and luminosity distributions have been the subject of numerous studies.
For pre--\textit{Swift} GRBs, these studies were based on the observed GRB redshift and peak flux distributions \citep{Firmani2004, Guetta2005}, on pseudo-redshifts \citep{Kocevski2006}, or on theoretical considerations.
The measure of hundreds of GRB redshifts with \textit{Swift} \citep{Gehrels2009} gave a new impulse to these studies, leading to better constraints on  the shape of the luminosity and energy distributions, their evolution with redshift and the role of low luminosity GRBs (see Table \ref{tab_LF} for a list of recent works).
However, despite this strong interest for the general shape of the GRB luminosity function, the question of the maximum GRB luminosity or their maximum energy is rarely discussed, probably because the most energetic GRBs (with \eiso\ $\approx 10^{54}$ erg) are very rare events.

This papers discusses the existence of a limit on the isotropic energy radiated by GRBs. 
In Section \ref{sec_eiso}, we construct the observed energy distribution of two samples of bright GRBs with well measured redshifts and spectral parameters. 
In Section \ref{sec_energetic}, the observed energy distribution is compared with theoretical distributions with or without a cutoff at high energies, and we show that the data strongly suggest the existence of a limit to the $\gamma$-ray isotropic energy radiated by GRBs. 
The significance and interpretation of this limit are discussed in Section \ref{sec_discussion}.

In this paper we use a flat cosmological model with H$_0$~=~{70 km s$^{-1}$ Mpc$^{-1}$} and $\Omega_{\rm M}$~=~0.3.

\begin{deluxetable}{l c c c c c c c}
\tablecaption{Models of GRB luminosity function. This table summarizes how the bright end of the GRB luminosity function has been parametrized in recent works. The slope refers to the high luminosity index for broken power law models, and to the slope below the cutoff luminosity for cutoff power law models. When it is mentioned, L$_{\rm max}$ indicates the maximum luminosity considered in the study. $\delta_n$ is the index of the density evolution and $\delta_l$ the index of luminosity evolution described in Section \ref{sub_eiso}.
\label{tab_LF}}
\tablehead{\colhead{Reference} & \colhead{Model}	&  \colhead{Slope}	&  \colhead{L$_{\rm br}$} 	&  \colhead{L$_{\rm cut}$}	& \colhead{L$_{\rm max}$} &  \colhead{$\delta_n$} &  \colhead{$\delta_l$} \\
\colhead{} & \colhead{}	&  \colhead{}	&  \colhead{erg~s$^{-1}$} 	&  \colhead{erg~s$^{-1}$}	& \colhead{erg~s$^{-1}$} &  \colhead{} &  \colhead{} }
%\colhead{} & \colhead{}	&  \colhead{}	&  \colhead{} 	&  \colhead{}	& \colhead{} & \colhead{index, $\delta_n$} &  \colhead{index, $\delta_l$} }
\startdata
\cite{Daigne2006} (SFR$_2$) & simple PL & $1.6$ & & & $4 \times 10^{53}$ & -- & -- \\
\cite{Salvaterra2007} & cutoff PL & $3.5$ & & $9.5 \times 10^{51}$ & & -- & -- \\
 & cutoff PL & $2.2$ & & $0.8 \times 10^{51}$ & & -- & 1.4 \\
\cite{Zitouni2008} (SFR$_2$) & broken PL & $2.0$ & $3 \times 10^{51}$ & & $3 \times 10^{53}$ & -- & -- \\
\cite{Dai2009} & broken PL & $1.3$ & $5 \times 10^{48}$ &  & & -- & -- \\
%\cite{Salvaterra2009} & Cutoff PL & $2.2$ & & $0.7 \times 10^{51}$ & & -- & 1.5 \\
\cite{Butler2010} & broken PL & $3$ & $5 \times 10^{52}$ &  & & B10\tablenotemark{a} & -- \\
\cite{Wanderman2010} & broken PL & $1.4$ & $3 \times 10^{52}$ & & & W10\tablenotemark{b} & -- \\
\cite{Salvaterra2012} & broken PL & $2.3$ & $3.8 \times 10^{52}$ & & & 1.7 & -- \\
 & cutoff PL & $2.1$ & & $3.1 \times 10^{51}$ & & 1.6 & -- \\
 & broken PL & $1.9$ & $0.6 \times 10^{51}$ & & & -- & 2.1 \\
 & cutoff PL & $2.0$ & & $0.2 \times 10^{51}$ & & -- & 2.3 \\
\cite{Shahmoradi2013}\tablenotemark{c} & log-normal & &  & $2.2 \times 10^{51}$ & & -- & -- \\
\cite{Howell2014} & broken PL & $2.6$ & $0.8 \times 10^{52}$ & & & W10 & -- \\
\cite{Lien2014} & broken PL & $3.0$ & $1.1 \times 10^{52}$ &  & & W10 & -- \\
\cite{Pescalli2015} & broken PL & $1.8$ & $2.8 \times 10^{51}$ & & & -- & 2.5 \\
\cite{Petrosian2015} &  broken PL & $3.2$ & $1 \times 10^{51}$ & & &-- & 2.3\\
 & cutoff PL & $0.5$ & & $1.4 \times 10^{51}$ & & -- & 2.3 \\
\cite{Tan2015} (RGRB2) &  broken PL & $2.4$ & $3.9 \times 10^{51}$ &  & & W10 & -- \\
&  broken PL & $2.1$ & $1.4 \times 10^{51}$ & & & W10 & 0.8 \\
\cite{Deng2016} &  broken PL & $2.5$ & $1.7 \times 10^{51}$ & & & -- & 1.14 \\
\enddata
\tablenotetext{a}{\cite{Butler2010} propose a parametrization of the GRB formation rate which cannot be represented by a simple index $\delta_n$. We note B10 this parametrization, which predicts an excess of GRBs over the SFR of \cite{Hopkins2006}, by a factor $\sim$3.7 at redshift z=5. }
\tablenotetext{b}{\cite{Wanderman2010} propose a parametrization of the GRB formation rate, which cannot be represented by a simple index $\delta_n$. We note W10 this parametrization, which predicts an excess of GRBs over the SFR of \cite{Hopkins2006}, by a factor $\sim$3.4 at redshift z=5. Various other studies use this parametrization. }
\tablenotetext{c}{L$_{\rm cut}$ gives the center of the log-normal distribution, the width of the distribution is: log($\sigma_{\rm L}$) = $-0.22$.}
\end{deluxetable}

%________________________________________________________________

\section{The GRB isotropic energy distribution}
\label{sec_eiso}

%________________________________________________________________
%________________________________________________________________

\subsection{Construction of two GRB samples}
\label{sub_sample}
For the purpose of this study, we construct two samples of long GRBs with reliable redshifts, fluence spectral parameters  and homogeneous selection criteria.
These samples are based on GRBs detected with \textit{Fermi}/GBM \citep{Meegan2009} and Konus\textit{--Wind} (KW, \citealt{Aptekar1995}).
These two instruments measure the spectral parameters of the prompt emission over a broad energy range, allowing reliable calculations of \eiso . 
For each instrument, we select long GRBs (with \dur $>$2~s) according to the following criteria: a peak flux large enough to avoid detection threshold effects, a duration shorter than 1000 seconds and a best fit spectral model which is curved (i.e. not a simple power law).
The isotropic energy release \eiso\ is calculated in the energy range $[1 - 10^4]$ keV in the cosmological rest-frame, following a standard procedure: we first compute the bolometric fluence in the energy range $[{1 \over 1+z} - {10^4 \over 1+z}]$ keV from the best fit fluence spectral model according to equation \ref{eq_sbol}, then we compute \eiso\ from the bolometric fluence according to equation \ref{eq_eiso}. 
N(E) in equation \ref{eq_sbol} is the photon spectrum of the GRB, which is obtained from the best fit fluence spectrum in the \textit{Fermi} GBM Burst Catalog for \textit{Fermi} GRBs \citep{Gruber2014,vonKienlin2014} and from the catalog of Konus\textit{--Wind} bursts with known redshifts for Konus GRBs \citep{Tsvetkova2017}.
The spectral parameters are listed in Tables \ref{tab_fermi} and \ref{tab_konus}.
The ratio of integrals in equation \ref{eq_sbol}, is the k-correction \citep{Bloom2001}, which is also listed in Tables \ref{tab_fermi} and \ref{tab_konus}. 

\begin{equation}
\label{eq_sbol}
S_{bol} = S_\gamma~{\int^{10^{4} \over 1+z}_{1 \over 1+z} E~N(E)~dE \over \int^{E_{\rm max}}_{E_{\rm min}} E~N(E)~dE} 
\end{equation}

\begin{equation}
\label{eq_eiso}
E_{\rm iso} = {4~\pi~D_l^2~S_{\rm bol} \over 1+z} 
\end{equation}

\begin{figure}[t]
\centering
\includegraphics[width=0.48\textwidth]{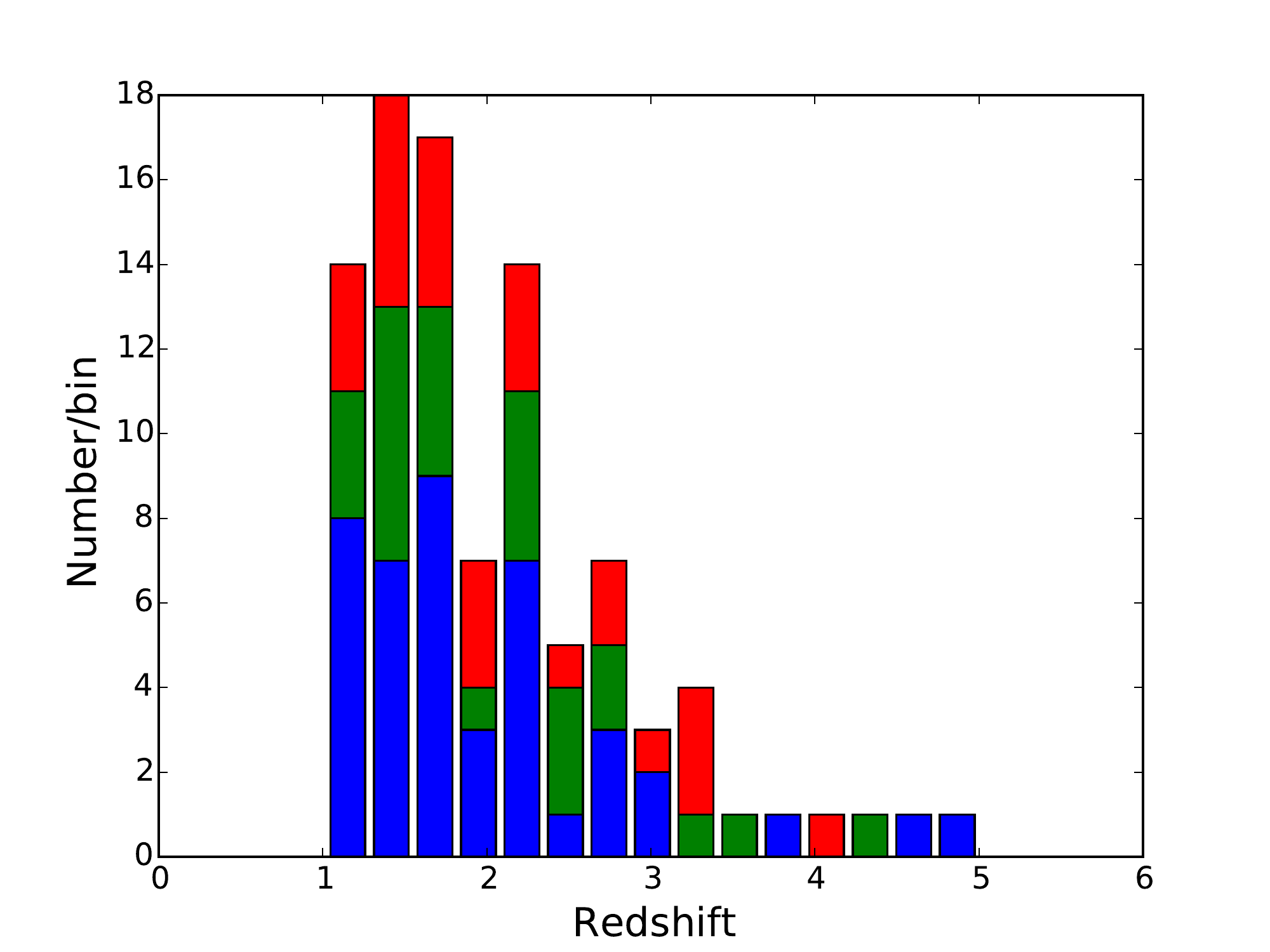}
\includegraphics[width=0.48\textwidth]{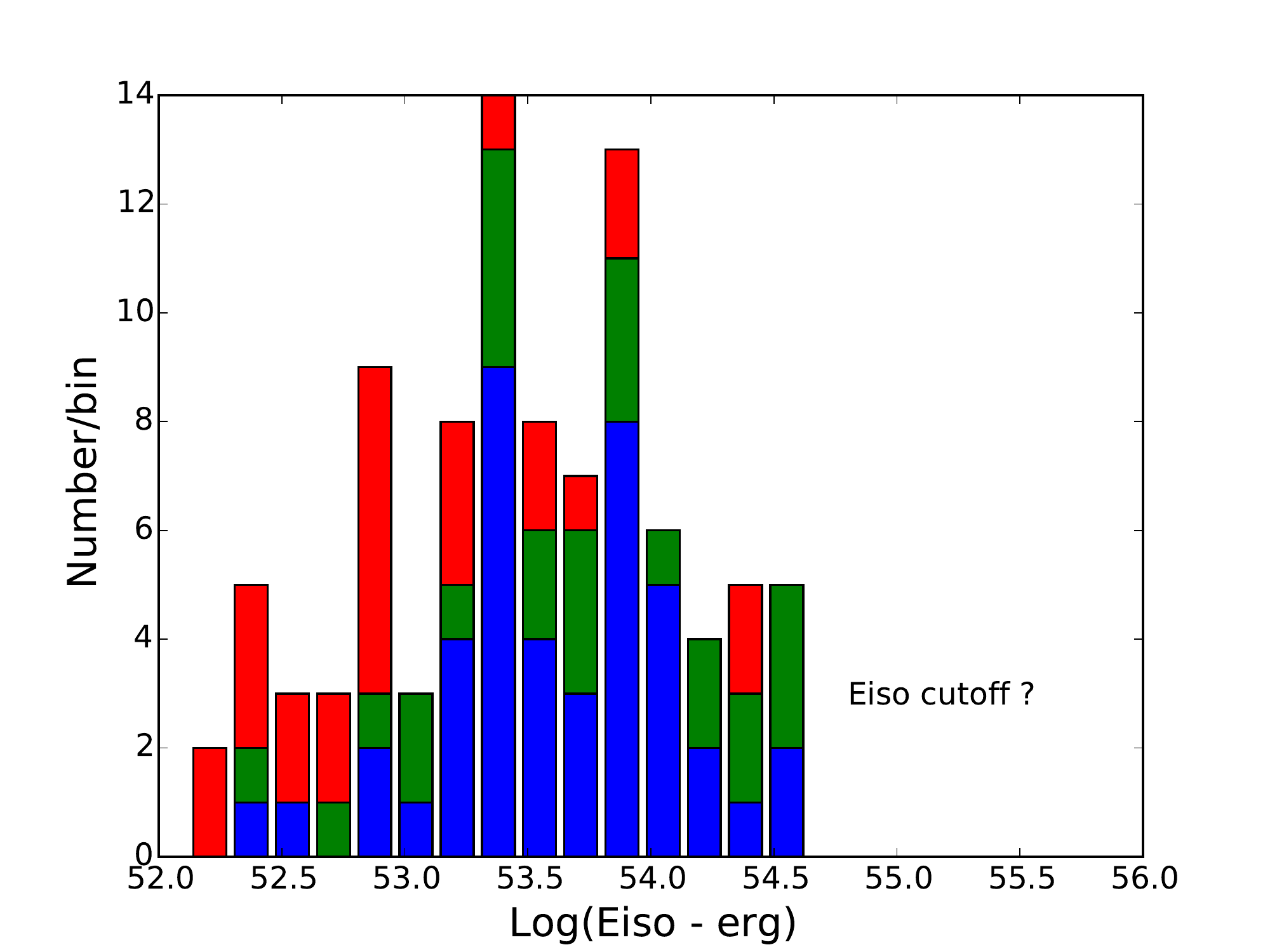}
\caption{Histograms of redshift (left) and \eiso\ (right) of GRBs in this study.
\textit{Fermi}-only GRBs are indicated in red, Konus-only GRBs in blue, and GRBs detected by \textit{Fermi} and Konus are indicated in green.}
\label{fig_histograms}
\end{figure}

Since we are mostly interested in energetic GRBs, which are rare in the local universe, we restrict our analysis to GRBs in the range $1 \le {\rm z} \le 5$. 
This cut has two advantages: it limits the impact of redshift evolution within our sample and it avoids the complex optical selection effects taking place when the Lyman alpha forest enters the R band channel at z $\ge 6$.
Moreover, since the volume enclosed within z=1 represents only 8\% of the volume enclosed within z=5 we keep 92\% of energetic GRBs, while removing from our sample low energy GRBs which are not useful for our analysis.
Figure \ref{fig_histograms} shows the distribution in redshift and \eiso\ of the GRBs in our sample.
%Therefore, the conclusions of this paper are only applicable to GRBs within this range of redshifts.
% Discuter l'obtention des redshifts

\begin{figure}[t]
\centering
\includegraphics[width=0.60\textwidth]{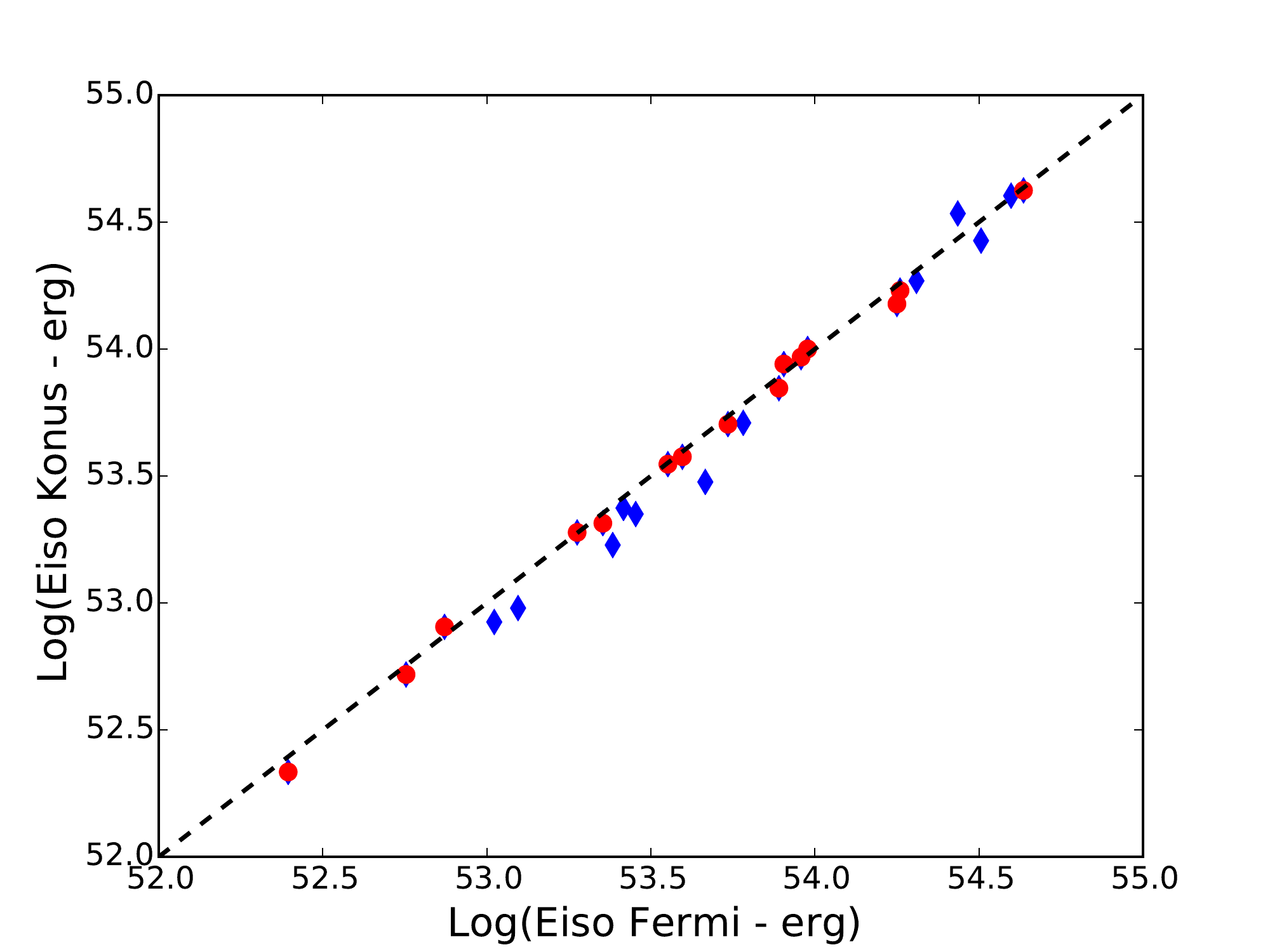}
\caption{Comparison of \eiso\ measured by \textit{Fermi}/GBM and Konus\textit{--Wind} for \fkngrb\ GRBs detected in common. 
The symbols refer to the model used for the spectral fit: red circles correspond to GRBs fitted with the same spectral model in Konus\textit{--Wind} and \textit{Fermi}/GBM, and blue diamonds to GRBs fitted with different spectral models. 
The 90\% error bars have typically the size of the symbols.
The dashed line indicates the equality of \eiso\ measured with \textit{Fermi}/GBM and Konus\textit{--Wind}.}
\label{fig_KFcomp}
\end{figure}

\subsubsection{Peak flux and redshift selection}
\label{ssub_selection}
GRB samples in this study are subject to two selection effects: in peak flux and in redshift, the construction of a reliable energy distribution is only possible if we correctly take into account the impact of these selections.
Considering the selection of GRBs with a redshift, it has been shown by \cite{Turpin2016} that GRBs with small and large afterglow optical fluxes have similar distributions of \epi\ (the maximum of the $\nu$F$\nu$ fluence spectrum), \eiso\ and \liso\ (the isotropic equivalent luminosity).
These authors conclude that the rest-frame distributions of  \epi , \eiso\ and \liso\ are not significantly distorted when they are computed from GRBs with a redshift.
%This is especially true for energetic GRBs with \eiso $\ge 10^{53}$ erg used in this study. 
We thus consider for the sake of this study that we do not bias the bright end of the GRB energy distribution when we study the distribution of GRBs with a redshift. 

Considering the impact of peak flux selection, we construct GRB samples with a peak flux threshold in the trigger energy range that is typically 50\% higher than the trigger threshold. 
This procedure transforms the complex detection instrument threshold into a well-defined \textit{sample threshold}, at the expense of loosing the faintest GRBs. 
The chosen values ensure that GRBs in our samples will be detected in most observing conditions.
In the rest of this paper, we use the sample threshold to evaluate the impact of peak flux selection effects.
% Ajouter un mot sur la mesure des redshifts et sur les effets de sŽlection qu'ils apportent, citer ??? en plus de Turpin

\subsubsection{The \textit{Fermi}/GBM sample}
\label{ssub_fermi}
We construct the \textit{Fermi}/GBM sample from the list of GRBs with a redshift provided in the online GRB table of Greiner\footnote{http://www.mpe.mpg.de/$\sim$jcg/grbgen.html}, from August 2008 to mid-2016. 
The best fit spectral model is extracted from the \textit{Fermi} GBM Burst Catalog \citep{Gruber2014,vonKienlin2014}.
In a first cut, we select GRBs with a 1-second peak flux larger than \Pf~=~1.05 ph cm$^{-2}$ s$^{-1}$ in the energy range [50--300] keV. 
This is 1.5 times larger than the detection threshold of 0.7 ph cm$^{-2}$ s$^{-1}$ \citep{NarayanaBhat2016}.
The requirement for a curved energy spectrum eliminates 6 GRBs whose best fit fluence spectrum is a power law.
The duration cut eliminates one very long GRB (GRB~091024).
After these cuts, we are left with a list of \fngrb\ GRBs given in Table~\ref{tab_fermi}.

The median 1-second peak flux of GRBs in our sample is {\Pf\ = 2.45 ph cm$^{-2}$ s$^{-1}$} in the 50-300 keV energy range, and the median redshift z=\fmedz , which is smaller than the median redshift of \textit{Swift} GRBs, z=2.2 \citep{Coward2013}.

\subsubsection{The Konus\textit{--Wind} sample}
\label{ssub_konus}
The Konus\textit{--Wind} instrument collects GRB spectral data since 1994 over a wide energy range ($\sim$10 keV -- 10 MeV, nominally). 
In the period from 1997 January to mid-2016, KW detected $\sim$150 GRBs with known redshifts in the triggered mode, of which 92 are in the range $1 \le {\rm z} \le 5$. 
For details of the KW analysis and for the complete catalog of the KW bursts with known redshifts see \cite{Tsvetkova2017}.
We select here GRBs which have a 1-second peak flux larger than \Pf~=~3.5 ph~cm$^{-2}$~s$^{-1}$ in the energy range [$50-200$] keV and a duration shorter than 1000 seconds.
The best fit fluence spectral model is chosen from the exponentially cutoff power-law (CPL) and the Band GRB function \citep{Band1993} based on the difference in $\chi^2$ between the fits. 
The criterion for accepting the Band function as the best-fit model is a  $\chi^2$ reduction of at least 6. 
We eliminate one GRB with a power law fluence spectrum.
After these cuts, we are left with a list of \kngrb\ GRBs given in Table~\ref{tab_konus}.
The median 1-second peak flux of GRBs in our sample is \Pf~=~7.3 ph~cm$^{-2}$~s$^{-1}$ in the [$50-200$] keV energy range, and the median redshift z=\kmedz, which is again smaller than the median redshift of \textit{Swift} GRBs, but comparable with the median redshift of the \textit{Fermi} sample.

\subsubsection{Comparison of the two samples}
\label{ssub_comparison}
The \textit{Fermi}/GBM and Konus\textit{--Wind} samples contain \fkngrb\ GRBs in common. 
Figure \ref{fig_KFcomp} compares \eiso\ measured with the \textit{Fermi}/GBM and Konus\textit{--Wind}.
It evidences few facts: the two measurements agree within 25\% for a large majority of GRBs (24/\fkngrb ) ; the agreement is better when the same model is used by the two instruments (red points) ; and in the few cases with a significant difference \textit{Fermi} measures larger \eiso\ as demonstrated by the location of the majority of the blue points below the dashed line.
The good agreement on \eiso\ measured with two instruments with different energy thresholds and different methods of spectral analyses indicate that the two samples used in this study are reliable, with no strong systematic uncertainties.

The largest differences reach 30\% for GRB~081222 (\eiso\ = $2.4 \times 10^{53}$ erg for \textit{Fermi}/GBM vs \eiso\ = $1.7 \times 10^{53}$ erg for Konus\textit{--Wind}), and 35\% for GRB~110731A (\eiso\ = $4.6 \times 10^{53}$ erg for \textit{Fermi}/GBM vs \eiso\ = $3.0 \times 10^{53}$ erg for Konus\textit{--Wind}).
We note that these differences reduce to 21\% and 30 \% respectively when the same model is used to fit the spectra from the two instruments.
These differences do not impact the analysis presented here, which is based on the number of GRBs found in broad classes of luminosity (0.5 dex, corresponding to a factor 3).

Three GRBs detected in common have durations that differ by more than a factor two between \textit{Fermi}/GBM and Konus\textit{--Wind} (GRB~081121, GRB~160509A, GRB~160625B), however the fluences measured by the two instruments differ by less than 5\% , validating the measure of \eiso .

%The slightly larger values measured with \textit{Fermi} might be explained by the impact of the BGO detectors of the \textit{Fermi}/GBM, which have a larger effective area at MeV energies and a higher upper energy limit (~40 MeV), resulting in an improved capability to measure the contribution of MeV photons.
%A confirmation of this explanation would however require detailed spectral analyses of simultaneously detected GRBs, which is beyond the scope of this paper. 

\begin{figure}[t]
\centering
\includegraphics[width=0.48\textwidth]{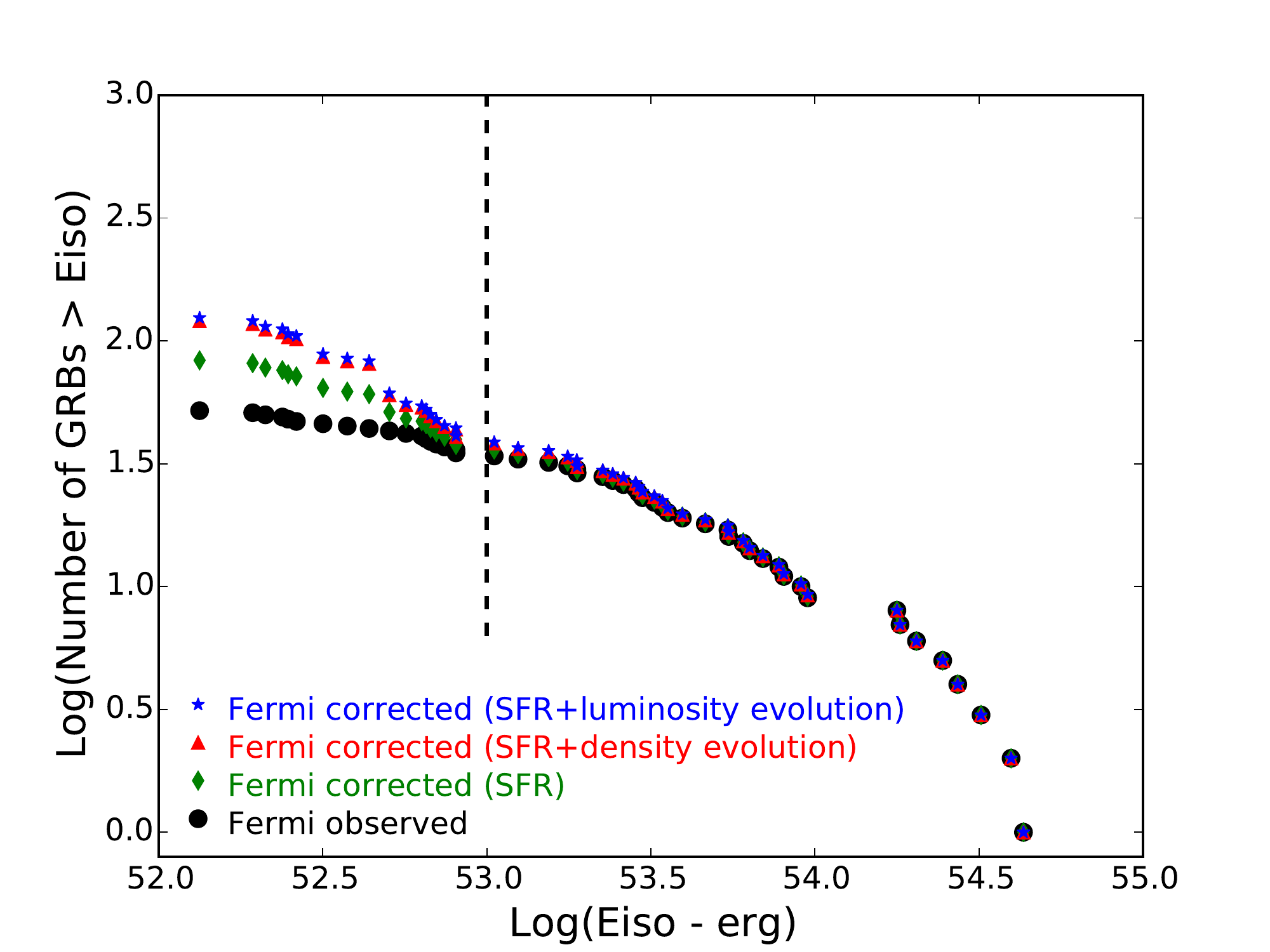}
\includegraphics[width=0.48\textwidth]{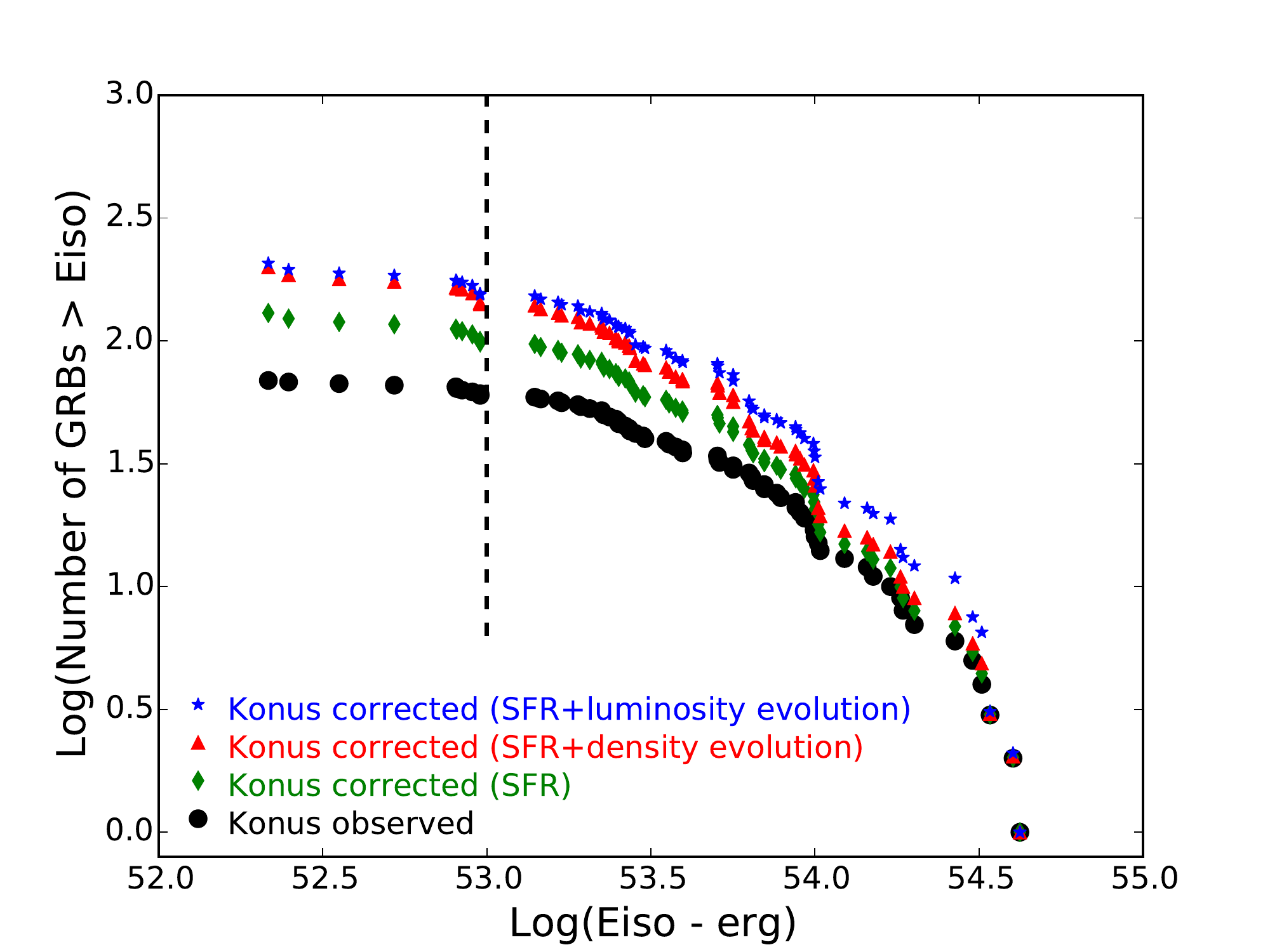}
\caption{Cumulative distribution of GRB isotropic energy of \fngrb\ GRBs detected by \textit{Fermi}/GBM (left panel) and \kngrb\ GRBs detected by Konus\textit{--Wind} (right panel). The black circles show the observed distribution, while green diamonds, red triangles and blue stars respectively show the corrected distribution assuming that GRBs follow the SFR of \cite{Hopkins2006}, the SFR multiplied by a density evolution proportional to \devolx , or the SFR with \eiso\ evolution proportional to \levolx . 
The smaller values of \zmax\ for Konus--\textit{Wind} lead to larger corrections on the right panel. 
The vertical dashed lines mark out GRBs with \eiso $\ge 10^{53}$ erg, which are studied in Section \ref{sec_energetic}.}
\label{fig_EisoCDF}
\end{figure}

\subsection{The corrected \eiso\ distribution}
\label{sub_eiso}
Figure \ref{fig_EisoCDF} shows with black circles the observed cumulative distribution of \eiso\ for \textit{Fermi}/GBM and Konus\textit{--Wind} GRBs.
These distributions do not represent the true GRB energy distribution since many GRBs are not detectable within the entire volume under study ($1 \le {\rm z} \le 5$).
In order to construct the true \eiso\ distribution, we use a two step procedure correcting for the detection inefficiency of GRBs in our samples.
First, we compute for each GRB the maximum distance at which its peak flux stays above the peak flux threshold of the sample. 
This ``GRB horizon'', \zmax\ is given in Tables \ref{tab_fermi} and \ref{tab_konus}. 
For bright GRBs visible to distances larger than z~=~5, we set the horizon to z~=~5, which is the redshift limit of our samples.
Tables \ref{tab_fermi} and \ref{tab_konus} allow comparing the "horizon" of \textit{Fermi}/GBM and Konus\textit{--Wind} for GRBs detected in common.
Unlike \eiso , we find some differences here (e.g. GRB~120624B which is detectable to \textbf{\zmax = 2.92} with Konus\textit{--Wind} and to \zmax = 5.0 with \textit{Fermi}/GBM).
These differences are readily explained by the different sensitivities of the two instruments which led us to adopt different peak flux thresholds for the two samples, as explained in section \ref{ssub_selection}.
The calculation of \zmax\ permits taking this parameter into account in our analysis.

In a second step we compute a ``weight'' for each GRB, given by the ratio of the number of GRBs of this type within z~=~5 to the number of GRBs within \zmax .
With this method, bright GRBs visible out to z~=~5 will be given a weight of 1, while fainter GRBs with an horizon smaller than z~=~5 will be given a weight larger than 1.
The weight of a GRB is thus the inverse of its detection efficiency within the volume under study.

The calculation of these weights require a GRB world model, which describes the volume density evolution and the energy evolution of GRBs with redshift.
For the purpose of this paper, which aims at studying the bright end of the energy distribution in a restricted range of redshift, we have limited our analysis to three simple cases. 
First, a model with no evolution, where the number of GRBs is proportional to the Star Formation Rate (SFR) proposed by \cite{Hopkins2006} and \cite{Li2008}. 
This model is described by equations \ref{eq_sfr} and \ref{eq_nzmax0}, and it leads to the weights labeled W$_{\rm sfr}$.
Second, a model with density evolution described by equation \ref{eq_nzmaxd} with $\delta_n = 1.6$, leading to the weights labeled W$_{\rm d}$.
Third, a model with luminosity evolution described by equation \ref{eq_nzmaxl} with $\delta_l = 1.5$.
Apart from GRBs visible out to z=5, which have weight unity, the weights of other GRBs depend on the model.
For models with luminosity evolution, the weights also depend on the luminosity function, the weights labeled W$_{\rm pl}$ refer to GRBs with a power law energy function, while W$_{\rm cpl}$ refer to GRBs with a cutoff power law energy function.
The indices for the density and luminosity evolution are typical values inferred from recent studies \citep[e.g.][]{Kistler2008, Salvaterra2012, Howell2014, Petrosian2015, Tan2015}.
Equations \ref{eq_sfr}, \ref{eq_nzmax0}, \ref{eq_nzmaxd}, \ref{eq_nzmaxl} give the formulae used for the calculation of the Star Formation Rate (equations \ref{eq_sfr}) and for the calculation of the number of GRBs closer than redshift $z_a$, N($< z_a$), for three cases of redshift evolution (equations \ref{eq_nzmax0}, \ref{eq_nzmaxd}, \ref{eq_nzmaxl}).

\begin{equation}
\label{eq_sfr}
SFR(z) \propto {{0.0157+0.118 z} \over {1+({z \over3.23})^{4.66}}}
\end{equation}

\begin{equation}
\label{eq_nzmax0}
      N(<z_a) \propto  \int_{0}^{z_a} {SFR(z)} {1 \over  (1+z)} {dV(z) \over  dz}  dz  \quad - \quad  {\rm no\  evolution} 
\end{equation}

\begin{equation}
\label{eq_nzmaxd}
      N(<z_a) \propto  \int_{0}^{z_a} {SFR(z)} {1 \over  (1+z)} {dV(z) \over  dz} (1+z)^{\delta_n} dz  \quad - \quad {\rm density\  evolution}
\end{equation}

\begin{equation}
\label{eq_nzmaxl}
      N(<z_a) \propto  \int_{0}^{z_a} {SFR(z)}  {1 \over  (1+z)} {dV(z) \over  dz}  {\phi({L \over (1+z)^{\delta_l}}) \over \phi(L)} dz  \quad - \quad {\rm luminosity\  evolution}
\end{equation}

For the density evolution model, the GRB rate is multiplied by nearly a factor six from redshift one to redshift five compared to the no evolution scenario.
For the luminosity evolution model, the GRB energy increases by about a factor five from redshift one to redshift five. 
In this model, the increase of the GRB rate with the redshift depends on the energy of the GRBs and on the shape of the energy distribution. 

Tables \ref{tab_fermi} and \ref{tab_konus} give the weights of GRBs for the different GRB world models studied here.
They show that the models with density or luminosity evolution have larger weights and require larger corrections because the comoving GRB density increases faster with redshift, leading to a higher fraction of undetected bursts for the same \zmax .

Having the weight of each GRB in our sample, we can then compute the corrected \eiso\ distribution.
Figure \ref{fig_EisoCDF} shows with green diamonds the corrected distributions for a model with no evolution, with red triangles the corrected distributions for the model with density evolution, and with blue stars the corrected distributions for a model with luminosity evolution.
The left panel shows the energy distributions derived from \textit{Fermi}/GBM observations and the right panel those derived from (less sensitive) Konus\textit{--Wind} observations.
The corrected \eiso\ distributions, like the observed one, exhibit a break around \eiso~$= 1-3 \times 10^{54}$ erg, which is the topic of this paper.

\newpage
\begin{deluxetable}{l l l l l l l l l l l l}
\tabletypesize{\footnotesize}
\tablecaption{Table of \fngrb\ GRBs detected by \textit{Fermi}/GBM used in this study. 
The 12 columns give respectively the name of the GRB, its duration \dur , its fluence in the [$10-10^3$] keV energy range, the parameters of the fluence spectral model, the redshift and k-correction \citep{Bloom2001}, \eiso , \zmax , and the weights of the GRBs for the four models under study (see section \ref{sub_eiso}). 
The spectral parameters and the names of the spectral models are taken from the \textit{Fermi} GBM Burst Catalog \citep{Gruber2014,vonKienlin2014}.
For the COMP model, the two parameters in the table are \epeak\ in keV and the power law index.
For the BAND model, the three parameters in the table are \epeak\ in keV, the low energy power law index and the high energy power law index.
For the SBPL model, the three parameters in the table are the smoothly broken power law break energy in keV, the low energy power law index and the high energy power law index.
The errors on \eiso\ have been derived from the error on the fluence, according to equations \ref{eq_sbol} and \ref{eq_eiso}.
GRBs detected in common with Konus are indicated in bold.
\label{tab_fermi}}
\tablehead{\colhead{GRB}	& \colhead{T$_{90}$}& \colhead{S$_{\gamma}/10^{-7}$}& \colhead{Spectral model} & \colhead{z }& \colhead{k-cor.}&  \colhead{E$_{\rm iso}/10^{52}$}	&  \colhead{\zmax} 	&  \colhead{W$_{\rm sfr}$}	& \colhead{W$_{\rm d}$} &  \colhead{W$_{\rm pl}$} &  \colhead{W$_{\rm cpl}$} \\
&  \colhead{s} &  \colhead{erg.cm$^{-2}$} & & & & \colhead{erg} & & & & & }
\startdata
GRB080804972 &  24.7&    91& SBPL     109   -0.70   -1.93 &  2.20 & 1.48 & $ 15.4\pm0.30$ &  3.06 &  1.37 &  1.76 &  2.07 &  1.81 \\
GRB080810549 & 107.5&   108& COMP     855   -1.18 &  3.35 & 1.37 & $ 34.3\pm0.25$ &  3.39 &  1.23 &  1.48 &  1.68 &  1.53 \\
\bf GRB080916009 &  63.0&   603& SBPL     302   -1.14   -2.09 &  4.35 & 1.52 & $272.5\pm0.52$ &  5.00 &  1.00 &  1.00 &  1.00 &  1.00 \\
\bf GRB081121858 &  42.0&   153& BAND     161   -0.43   -2.09 &  2.51 & 1.52 & $ 28.4\pm0.68$ &  5.00 &  1.00 &  1.00 &  1.00 &  1.00 \\
\bf GRB081221681 &  29.7&   300& COMP      88   -0.91 &  2.26 & 1.52 & $ 39.4\pm0.19$ &  5.00 &  1.00 &  1.00 &  1.00 &  1.00 \\
\bf GRB081222204 &  18.9&   119& BAND     143   -0.86   -2.31 &  2.77 & 1.52 & $ 24.2\pm0.32$ &  5.00 &  1.00 &  1.00 &  1.00 &  1.00 \\
\bf GRB090102122 &  26.6&   279& COMP     417   -0.94 &  1.55 & 1.52 & $ 18.8\pm0.07$ &  4.15 &  1.07 &  1.15 &  1.22 &  1.16 \\
GRB090113778 &  17.4&    16& COMP     178   -1.28 &  1.75 & 1.12 & $  1.3\pm0.07$ &  2.26 &  2.21 &  3.49 &  4.65 &  3.58 \\
\bf GRB090323002 & 135.2&  1181& SBPL     345   -1.34   -2.27 &  3.57 & 1.52 & $396.2\pm0.96$ &  5.00 &  1.00 &  1.00 &  1.00 &  1.00 \\
GRB090516353 & 123.1&   172& COMP     164   -1.54 &  4.11 & 1.26 & $ 69.4\pm0.37$ &  4.37 &  1.05 &  1.10 &  1.14 &  1.12 \\
GRB090902462 &  19.3&  2218& SBPL    1170   -1.09   -4.85 &  1.82 & 1.36 & $245.3\pm0.58$ &  5.00 &  1.00 &  1.00 &  1.00 &  1.00 \\
\bf GRB090926181 &  13.8&  1466& SBPL     202   -0.98   -2.31 &  2.11 & 1.52 & $203.9\pm0.78$ &  5.00 &  1.00 &  1.00 &  1.00 &  1.00 \\
GRB090926914 &  55.6&   108& COMP      86    0.04 &  1.24 & 1.02 & $  4.4\pm0.10$ &  1.36 &  9.35 & 20.48 & 32.65 & 21.61 \\
GRB091020900 &  24.3&    83& COMP     244   -1.26 &  1.71 & 1.11 & $  6.7\pm0.20$ &  2.70 &  1.62 &  2.26 &  2.79 &  2.32 \\
GRB091208410 &  12.5&    62& COMP     127   -1.34 &  1.06 & 1.16 & $  2.1\pm0.11$ &  2.49 &  1.84 &  2.73 &  3.49 &  2.79 \\
\bf GRB100414097 &  26.5&   885& COMP     668   -0.63 &  1.37 & 1.52 & $ 54.3\pm0.19$ &  4.58 &  1.03 &  1.06 &  1.08 &  1.07 \\
GRB100615083 &  37.4&    87& COMP     144   -1.35 &  1.40 & 1.16 & $  5.0\pm0.08$ &  1.96 &  3.01 &  5.27 &  7.43 &  5.48 \\
\bf GRB100728095 & 165.4&  1279& BAND     290   -0.64   -2.70 &  1.57 & 1.52 & $ 95.0\pm0.71$ &  3.93 &  1.10 &  1.22 &  1.31 &  1.27 \\
GRB100728439 &  10.2&    33& COMP     160   -0.98 &  2.11 & 1.06 & $  3.8\pm0.12$ &  2.88 &  1.48 &  1.97 &  2.38 &  2.01 \\
GRB100814160 & 150.5&   149& COMP     156   -0.50 &  1.44 & 1.02 & $  8.1\pm0.08$ &  2.39 &  1.98 &  3.01 &  3.92 &  3.11 \\
\bf GRB100906576 & 110.6&   233& SBPL      27   -0.89   -1.86 &  1.73 & 1.52 & $ 26.1\pm0.11$ &  3.89 &  1.11 &  1.23 &  1.33 &  1.25 \\
GRB110213220 &  34.3&    94& COMP     113   -1.57 &  1.46 & 1.29 & $  6.5\pm0.06$ &  2.61 &  1.71 &  2.44 &  3.06 &  2.51 \\
\bf GRB110731465 &   7.5&   229& SBPL     287   -1.04   -2.96 &  2.83 & 1.52 & $ 46.3\pm0.19$ &  5.00 &  1.00 &  1.00 &  1.00 &  1.00 \\
\bf GRB120119170 &  55.3&   387& BAND     183   -0.96   -2.37 &  1.73 & 1.52 & $ 35.6\pm0.21$ &  4.46 &  1.04 &  1.08 &  1.12 &  1.09 \\
GRB120326056 &  11.8&    33& SBPL      31   -0.92   -2.40 &  1.80 & 1.22 & $  3.2\pm0.08$ &  2.27 &  2.18 &  3.44 &  4.57 &  3.54 \\
\bf GRB120624933 & 271.4&  1916& SBPL     358   -1.02   -2.23 &  2.20 & 1.52 & $320.9\pm0.55$ &  5.00 &  1.00 &  1.00 &  1.00 &  1.00 \\
\bf GRB120711115 &  44.0&  1943& BAND    1319   -0.98   -2.80 &  1.41 & 1.52 & $181.7\pm0.35$ &  5.00 &  1.00 &  1.00 &  1.00 &  1.00 \\
GRB120716712 & 237.1&   144& BAND      85   -0.76   -1.84 &  2.49 & 1.43 & $ 29.1\pm0.14$ &  4.24 &  1.06 &  1.13 &  1.18 &  1.14 \\
GRB120811649 &  14.3&    34& COMP      61   -0.93 &  2.67 & 1.16 & $  6.3\pm0.64$ &  3.18 &  1.31 &  1.64 &  1.90 &  1.66 \\
\bf GRB121128212 &  17.3&    93& SBPL      43   -0.91   -2.48 &  2.20 & 1.52 & $ 12.4\pm0.25$ &  4.78 &  1.01 &  1.03 &  1.04 &  1.03 \\
\bf GRB130518580 &  48.6&   946& BAND     398   -0.91   -2.25 &  2.49 & 1.52 & $177.8\pm0.48$ &  5.00 &  1.00 &  1.00 &  1.00 &  1.00 \\
GRB131011741 &  77.1&    89& COMP     274   -0.96 &  1.87 & 1.06 & $  8.1\pm0.10$ &  2.68 &  1.64 &  2.30 &  2.86 &  2.37 \\
GRB131105087 & 112.6&   238& COMP     266   -1.26 &  1.69 & 1.12 & $ 18.8\pm0.14$ &  3.04 &  1.38 &  1.77 &  2.09 &  1.83 \\
\bf GRB131108862 &  18.2&   357& SBPL     240   -1.04   -2.42 &  2.40 & 1.52 & $ 60.5\pm0.38$ &  5.00 &  1.00 &  1.00 &  1.00 &  1.00 \\
GRB140206304 &  27.3&   155& BAND     121    0.06   -2.35 &  2.73 & 1.16 & $ 29.8\pm0.24$ &  5.00 &  1.00 &  1.00 &  1.00 &  1.00 \\
\bf GRB140213807 &  18.6&   212& BAND      87   -1.13   -2.26 &  1.21 & 1.52 & $ 10.5\pm0.05$ &  2.90 &  1.47 &  1.95 &  2.34 &  2.00 \\
GRB140423356 &  95.2&   181& BAND     121   -0.58   -1.83 &  3.26 & 1.36 & $ 54.6\pm0.61$ &  3.84 &  1.12 &  1.25 &  1.35 &  1.29 \\
\bf GRB140508128 &  44.3&   614& BAND     263   -1.19   -2.36 &  1.03 & 1.52 & $ 22.6\pm0.07$ &  5.00 &  1.00 &  1.00 &  1.00 &  1.00 \\
GRB140620219 &  45.8&    61& COMP     127   -1.28 &  2.04 & 1.14 & $  7.0\pm0.12$ &  2.57 &  1.75 &  2.52 &  3.18 &  2.59 \\
GRB140703026 &  84.0&    76& COMP     221   -1.28 &  3.14 & 1.12 & $ 17.6\pm0.20$ &  4.23 &  1.06 &  1.13 &  1.18 &  1.14 \\
\bf GRB140801792 &   7.2&   124& COMP     121   -0.40 &  1.32 & 1.52 & $  5.7\pm0.03$ &  3.44 &  1.22 &  1.45 &  1.62 &  1.46 \\
\bf GRB140808038 &   4.5&    32& COMP     123   -0.47 &  3.29 & 1.52 & $  7.4\pm0.12$ &  5.00 &  1.00 &  1.00 &  1.00 &  1.00 \\
GRB140907672 &  35.8&    65& COMP     142   -1.03 &  1.21 & 1.08 & $  2.6\pm0.04$ &  1.45 &  7.39 & 15.66 & 24.53 & 16.38 \\
GRB141028455 &  31.5&   348& BAND     294   -0.84   -1.97 &  2.33 & 1.44 & $ 63.2\pm0.27$ &  5.00 &  1.00 &  1.00 &  1.00 &  1.00 \\
\bf GRB141220252 &   7.6&    53& COMP     178   -0.82 &  1.32 & 1.52 & $  2.5\pm0.03$ &  2.97 &  1.42 &  1.86 &  2.21 &  1.89 \\
GRB141221338 &  23.8&    41& COMP     182   -1.18 &  1.45 & 1.09 & $  2.4\pm0.06$ &  2.01 &  2.83 &  4.87 &  6.80 &  5.03 \\
GRB150301818 &  13.3&    31& COMP     226   -1.12 &  1.52 & 1.08 & $  1.9\pm0.03$ &  1.88 &  3.31 &  5.94 &  8.50 &  6.14 \\
\bf GRB150314205 &  10.7&   816& BAND     347   -0.68   -2.60 &  1.76 & 1.52 & $ 77.7\pm0.20$ &  5.00 &  1.00 &  1.00 &  1.00 &  1.00 \\
\bf GRB150403913 &  22.3&   547& BAND     429   -0.87   -2.11 &  2.06 & 1.52 & $ 80.4\pm0.13$ &  5.00 &  1.00 &  1.00 &  1.00 &  1.00 \\
\bf GRB160509374 & 369.7&  1790& BAND     355   -1.02   -2.23 &  1.17 & 1.52 & $ 90.7\pm0.13$ &  5.00 &  1.00 &  1.00 &  1.00 &  1.00 \\
\bf GRB160625945 & 454.7&  5692& BAND     649   -0.95   -2.37 &  1.41 & 1.52 & $432.2\pm1.19$ &  5.00 &  1.00 &  1.00 &  1.00 &  1.00 \\
GRB160629930 &  64.8&   131& COMP     291   -1.03 &  3.33 & 1.07 & $ 32.4\pm0.14$ &  5.00 &  1.00 &  1.00 &  1.00 &  1.00 \\
\enddata
\end{deluxetable}

\newpage
\begin{deluxetable}{l l l l l l l l l l l l}
\tabletypesize{\footnotesize}
\tablecaption{Table of \kngrb\ GRBs detected by Konus\textit{--Wind} used in this study. 
The 12 columns give respectively the name of the GRB, its duration \dur , its fluence in the [$10-10^4$] keV energy range, the parameters of the fluence spectral model, the redshift and k-correction \citep{Bloom2001}, \eiso , \zmax , and the weights of the GRBs for the four models under study (see section \ref{sub_eiso}). 
The GRB parameters have been extracted from the Konus-WIND catalog of GRBs with known redshifts \citep{Tsvetkova2017}.
For the COMP model, the two parameters in the table are \epeak\ in keV and the power law index.
For the BAND model, the three parameters in the table are \epeak\ in keV, the low energy power law index and the high energy power law index.
The errors on \eiso\ have been derived from the error on the fluence, according to equations \ref{eq_sbol} and \ref{eq_eiso}.
GRBs detected in common with \textit{Fermi}/GBM are indicated in bold.
\label{tab_konus}}
\tablehead{\colhead{GRB}	& \colhead{T$_{90}$}& \colhead{S$_{\gamma}/10^{-7}$}& \colhead{Spectral model} & \colhead{z }& \colhead{k-cor.}&  \colhead{E$_{\rm iso}/10^{52}$}	&  \colhead{\zmax} 	&  \colhead{W$_{\rm sfr}$}	& \colhead{W$_{\rm d}$} &  \colhead{W$_{\rm pl}$} &  \colhead{W$_{\rm cpl}$} \\
&  \colhead{s} &  \colhead{erg.cm$^{-2}$} & & & & \colhead{erg} & & & & & }
\startdata
GRB990123 &  62.0&  2320& COMP     724   -0.94 &  1.60 & 1.35 & $201.0\pm8.50$ &  4.01 &  1.09 &  1.19 &  1.19 &  1.23 \\
GRB990506 & 128.6&  1600& BAND     296   -1.19   -2.09 &  1.31 & 1.47 & $103.8\pm5.75$ &  2.58 &  1.74 &  2.50 &  2.43 &  2.65 \\
GRB990510 &  55.9&   216& COMP     136   -1.35 &  1.62 & 1.16 & $ 16.5\pm1.12$ &  2.25 &  2.22 &  3.52 &  3.40 &  3.46 \\
GRB991216 &  14.5&  1956& BAND     353   -1.20   -2.23 &  1.02 & 1.44 & $ 76.5\pm1.60$ &  3.67 &  1.15 &  1.32 &  1.31 &  1.33 \\
GRB000131 &  96.5&   337& BAND     133   -0.90   -2.26 &  4.50 & 1.16 & $144.3\pm6.99$ &  5.00 &  1.00 &  1.00 &  1.00 &  1.00 \\
GRB000418 &  27.8&   218& COMP     116   -1.56 &  1.12 & 1.27 & $  9.0\pm0.60$ &  1.51 &  6.32 & 13.06 & 12.42 & 12.63 \\
GRB000911 &  23.4&  1071& BAND    1083   -0.82   -2.75 &  1.06 & 1.80 & $ 56.4\pm2.93$ &  1.64 &  4.82 &  9.48 &  9.04 &  9.99 \\
GRB000926 &  54.7&   209& COMP     108   -1.51 &  2.04 & 1.27 & $ 26.4\pm1.21$ &  2.44 &  1.91 &  2.86 &  2.78 &  2.84 \\
GRB010222 &  89.8&  1154& BAND     285   -1.26   -2.17 &  1.48 & 1.41 & $ 90.1\pm4.06$ &  2.95 &  1.43 &  1.88 &  1.85 &  1.95 \\
GRB020813 &  89.4&  1191& BAND     227   -0.90   -2.24 &  1.25 & 1.35 & $ 64.8\pm6.74$ &  2.58 &  1.73 &  2.50 &  2.43 &  2.56 \\
GRB050401 &  33.1&   182& BAND     105   -0.82   -2.31 &  2.90 & 1.19 & $ 39.5\pm2.74$ &  3.72 &  1.14 &  1.30 &  1.29 &  1.30 \\
GRB050603 &  11.2&   265& BAND     239   -0.69   -1.94 &  2.82 & 1.39 & $ 64.1\pm6.49$ &  5.00 &  1.00 &  1.00 &  1.00 &  1.00 \\
GRB051008 & 208.8&   385& COMP     550   -0.98 &  2.77 & 1.21 & $ 78.7\pm11.97$ &  3.25 &  1.28 &  1.58 &  1.56 &  1.61 \\
GRB060124 &  78.1&   202& COMP     239   -1.17 &  2.30 & 1.09 & $ 27.1\pm2.31$ &  2.88 &  1.48 &  1.97 &  1.93 &  1.96 \\
GRB061007 &  57.6&  1863& BAND     399   -0.70   -2.61 &  1.26 & 1.30 & $ 99.0\pm5.95$ &  2.74 &  1.58 &  2.18 &  2.13 &  2.28 \\
GRB061121 &  17.8&   486& COMP     607   -1.32 &  1.31 & 1.32 & $ 28.4\pm1.76$ &  2.90 &  1.46 &  1.94 &  1.90 &  1.93 \\
GRB061222 &  60.2&   225& COMP     298   -0.89 &  2.09 & 1.06 & $ 24.7\pm1.06$ &  2.85 &  1.50 &  2.01 &  1.97 &  2.00 \\
GRB070125 & 124.2&  1146& BAND     372   -1.10   -2.09 &  1.55 & 1.48 & $102.4\pm8.42$ &  3.26 &  1.28 &  1.57 &  1.55 &  1.62 \\
GRB070328 &  53.8&   370& BAND     386   -0.80   -2.00 &  2.06 & 1.50 & $ 56.3\pm10.37$ &  2.21 &  2.30 &  3.70 &  3.57 &  3.81 \\
GRB070521 &  31.8&   186& COMP     218   -0.92 &  1.70 & 1.05 & $ 14.0\pm0.71$ &  2.03 &  2.78 &  4.74 &  4.57 &  4.65 \\
GRB071003 &  21.4&   396& COMP     801   -0.97 &  1.60 & 1.41 & $ 35.9\pm2.86$ &  2.24 &  2.24 &  3.57 &  3.46 &  3.60 \\
GRB071020 &   2.7&    71& COMP     322   -0.65 &  2.15 & 1.04 & $  8.1\pm1.15$ &  2.98 &  1.41 &  1.84 &  1.80 &  1.81 \\
GRB071117 &   2.3&    63& COMP     278   -1.53 &  1.33 & 1.24 & $  3.5\pm0.36$ &  2.11 &  2.54 &  4.21 &  4.07 &  4.08 \\
GRB080319 &  10.2&   121& COMP     632   -1.21 &  1.95 & 1.31 & $ 14.6\pm2.72$ &  2.09 &  2.59 &  4.33 &  4.18 &  4.25 \\
GRB080411 &  42.8&   660& COMP     266   -1.52 &  1.03 & 1.23 & $ 22.5\pm1.14$ &  2.15 &  2.44 &  4.00 &  3.87 &  3.97 \\
GRB080514 &   5.7&   262& BAND     196   -0.53   -2.46 &  1.80 & 1.21 & $ 25.2\pm1.87$ &  4.19 &  1.07 &  1.14 &  1.14 &  1.14 \\
GRB080603 &  12.6&    51& COMP     101   -1.21 &  2.69 & 1.15 & $  9.5\pm2.33$ &  2.90 &  1.46 &  1.94 &  1.90 &  1.91 \\
GRB080605 &  13.7&   323& COMP     260   -0.89 &  1.64 & 1.05 & $ 22.7\pm0.73$ &  3.51 &  1.20 &  1.41 &  1.39 &  1.40 \\
GRB080607 &  28.7&   766& BAND     334   -0.71   -2.52 &  3.04 & 1.21 & $182.4\pm8.04$ &  5.00 &  1.00 &  1.00 &  1.00 &  1.00 \\
GRB080721 &  19.7&   625& BAND     490   -0.93   -2.45 &  2.59 & 1.31 & $123.1\pm8.53$ &  5.00 &  1.00 &  1.00 &  1.00 &  1.00 \\
\bf GRB080916 &  61.3&   788& BAND     505   -1.04   -2.26 &  4.35 & 1.24 & $341.8\pm45.47$ &  5.00 &  1.00 &  1.00 &  1.00 &  1.00 \\
\bf GRB081121 &  19.4&   151& COMP     254   -0.79 &  2.51 & 1.04 & $ 22.4\pm1.62$ &  2.59 &  1.72 &  2.47 &  2.40 &  2.45 \\
\bf GRB081221 &  29.2&   278& COMP      81   -1.03 &  2.26 & 1.13 & $ 37.6\pm1.22$ &  3.21 &  1.30 &  1.61 &  1.58 &  1.61 \\
\bf GRB081222 &  12.0&    96& COMP     192   -0.84 &  2.77 & 1.04 & $ 16.9\pm1.92$ &  3.13 &  1.33 &  1.68 &  1.65 &  1.67 \\
\bf GRB090102 &  15.3&   279& COMP     432   -0.90 &  1.55 & 1.12 & $ 19.0\pm1.65$ &  1.86 &  3.40 &  6.16 &  5.91 &  6.07 \\
GRB090201 &  67.3&   730& BAND     156   -0.90   -2.71 &  2.10 & 1.14 & $ 87.5\pm4.04$ &  3.68 &  1.15 &  1.32 &  1.30 &  1.34 \\
\bf GRB090323 & 133.0&  1187& BAND     417   -0.96   -2.10 &  3.60 & 1.30 & $401.7\pm51.76$ &  4.72 &  1.02 &  1.04 &  1.04 &  1.05 \\
GRB090709 &  77.3&   755& COMP     277   -0.86 &  1.80 & 1.05 & $ 63.0\pm1.94$ &  2.43 &  1.93 &  2.89 &  2.81 &  2.98 \\
\bf GRB090926 &  13.2&  1438& BAND     327   -0.79   -2.61 &  2.11 & 1.22 & $185.6\pm7.70$ &  5.00 &  1.00 &  1.00 &  1.00 &  1.00 \\
\bf GRB100414 &  21.7&   888& COMP     571   -0.49 &  1.37 & 1.19 & $ 50.6\pm1.68$ &  2.09 &  2.59 &  4.32 &  4.17 &  4.43 \\
GRB100606 &  59.1&   306& COMP     874   -1.00 &  1.55 & 1.47 & $ 27.3\pm3.48$ &  1.59 &  5.35 & 10.73 & 10.22 & 10.74 \\
\bf GRB100728 & 159.9&  1270& BAND     305   -0.65   -2.48 &  1.57 & 1.27 & $100.1\pm10.21$ &  2.01 &  2.81 &  4.82 &  4.64 &  5.30 \\
\bf GRB100906 &  90.1&   249& COMP     195   -1.60 &  1.73 & 1.28 & $ 23.6\pm4.29$ &  2.00 &  2.86 &  4.94 &  4.75 &  4.90 \\
GRB110422 &  22.3&   844& BAND     155   -0.70   -3.21 &  1.77 & 1.08 & $ 70.1\pm1.73$ &  4.36 &  1.05 &  1.10 &  1.10 &  1.11 \\
GRB110503 &   6.7&   253& BAND     220   -0.98   -2.71 &  1.61 & 1.18 & $ 19.3\pm1.43$ &  3.51 &  1.20 &  1.41 &  1.39 &  1.40 \\
\bf GRB110731 &   6.7&   164& COMP     288   -0.74 &  2.83 & 1.04 & $ 30.0\pm1.89$ &  4.05 &  1.08 &  1.18 &  1.17 &  1.18 \\
GRB111008 &  12.7&    70& COMP     104   -1.53 &  5.00 & 1.30 & $ 39.5\pm8.83$ &  5.00 &  1.00 &  1.00 &  1.00 &  1.00 \\
\bf GRB120119 &  39.6&   383& BAND     153   -0.85   -2.34 &  1.73 & 1.24 & $ 35.2\pm4.15$ &  2.44 &  1.91 &  2.87 &  2.79 &  2.88 \\
\bf GRB120624 & 267.9&  1911& COMP     560   -1.04 &  2.20 & 1.23 & $267.3\pm18.69$ &  2.92 &  1.45 &  1.92 &  1.88 &  2.22 \\
\bf GRB120711 &  41.3&  1979& BAND    1061   -0.97   -2.71 &  1.41 & 1.69 & $169.9\pm6.49$ &  2.43 &  1.92 &  2.88 &  2.80 &  3.28 \\
\bf GRB121128 &  10.0&    74& COMP      77   -0.99 &  2.20 & 1.13 & $  9.5\pm0.46$ &  3.03 &  1.38 &  1.78 &  1.75 &  1.76 \\
GRB130408 &   4.2&    75& BAND     271   -0.70   -2.30 &  3.76 & 1.20 & $ 25.2\pm8.80$ &  4.32 &  1.05 &  1.11 &  1.11 &  1.11 \\
GRB130505 &  14.7&  1580& BAND     593   -0.49   -2.04 &  2.27 & 1.59 & $302.6\pm11.52$ &  5.00 &  1.00 &  1.00 &  1.00 &  1.00 \\
\bf GRB130518 &  28.6&   740& BAND     332   -0.88   -1.96 &  2.49 & 1.44 & $150.6\pm15.98$ &  5.00 &  1.00 &  1.00 &  1.00 &  1.00 \\
GRB130701 &   3.7&    63& COMP      89   -1.10 &  1.16 & 1.13 & $  2.5\pm0.12$ &  1.78 &  3.85 &  7.19 &  6.88 &  6.91 \\
GRB130907 & 180.3&  5594& BAND     387   -0.90   -2.22 &  1.24 & 1.45 & $322.6\pm17.98$ &  2.99 &  1.41 &  1.83 &  1.80 &  2.16 \\
GRB131030 &  15.7&   647& BAND     196   -0.52   -3.05 &  1.29 & 1.10 & $ 30.3\pm1.98$ &  2.93 &  1.45 &  1.91 &  1.87 &  1.90 \\
\bf GRB131108 &  17.7&   343& COMP     358   -1.16 &  2.40 & 1.13 & $ 51.2\pm3.83$ &  3.63 &  1.16 &  1.34 &  1.33 &  1.35 \\
\bf GRB140213 &  16.4&   183& COMP     100   -1.40 &  1.21 & 1.21 & $  8.4\pm0.29$ &  1.91 &  3.18 &  5.65 &  5.43 &  5.49 \\
\bf GRB140508 & 149.7&   594& BAND     220   -1.17   -2.54 &  1.03 & 1.25 & $ 20.6\pm2.34$ &  2.85 &  1.50 &  2.02 &  1.98 &  2.00 \\
\bf GRB140801 &   6.2&   113& COMP     108   -0.44 &  1.32 & 1.03 & $  5.2\pm0.19$ &  1.65 &  4.72 &  9.24 &  8.82 &  8.90 \\
\bf GRB140808 &   5.2&    33& COMP     125   -0.94 &  3.29 & 1.07 & $  8.0\pm0.76$ &  3.47 &  1.21 &  1.43 &  1.41 &  1.41 \\
\bf GRB141220 &   6.9&    47& COMP     139   -0.55 &  1.32 & 1.03 & $  2.2\pm0.18$ &  1.49 &  6.64 & 13.84 & 13.16 & 13.21 \\
GRB150206 &  35.1&   370& BAND     228   -0.73   -2.20 &  2.09 & 1.31 & $ 50.5\pm6.15$ &  3.18 &  1.31 &  1.64 &  1.62 &  1.65 \\
\bf GRB150314 &  10.2&   775& BAND     350   -0.78   -2.90 &  1.76 & 1.18 & $ 70.1\pm3.25$ &  5.00 &  1.00 &  1.00 &  1.00 &  1.00 \\
\bf GRB150403 &  21.0&   595& BAND     373   -0.93   -2.06 &  2.06 & 1.45 & $ 87.3\pm7.74$ &  4.43 &  1.04 &  1.09 &  1.08 &  1.09 \\
GRB151021 &  57.0&   650& BAND     170   -1.14   -2.46 &  2.33 & 1.22 & $ 99.5\pm14.22$ &  3.05 &  1.38 &  1.77 &  1.74 &  1.83 \\
\bf GRB160509 &  28.5&  1749& BAND     288   -0.99   -2.08 &  1.17 & 1.50 & $ 92.9\pm14.02$ &  3.17 &  1.32 &  1.65 &  1.62 &  1.69 \\
\bf GRB160625 &  21.0&  5473& BAND     571   -0.80   -2.28 &  1.41 & 1.52 & $421.5\pm8.49$ &  5.00 &  1.00 &  1.00 &  1.00 &  1.00 \\
\enddata
\end{deluxetable}

%________________________________________________________________
\section{The most energetic GRBs}
\label{sec_energetic}
In this section we compare the distribution of  \eiso\ derived above with two models of the energy function: a simple power law (more correctly called the Pareto distribution) and a power law with a high energy cutoff (more correctly called the gamma distribution). 
Our goal is to assess the significance of the energy cutoff observed in figure \ref{fig_EisoCDF}.

Combining these two energy functions with the three GRB world models previously discussed (SFR, SFR+density evolution, SFR+luminosity evolution), we obtain a total of six models, that are compared with the data thanks to a chisquare test.
For the purpose of the test, we classify GRBs into 5 classes of \eiso\ ranging from 10$^{53}$ to 10$^{56}$ erg.
Within each class of \eiso\ we compute the number of GRBs predicted by the theoretical model, taking into account a detection efficiency defined as the average weight of GRBs in this class, and we compare the theoretical numbers with the observed numbers.

The comparison involves the normalization of the theoretical numbers to the number of observed GRBs with energies larger than $10^{53}$ erg: \kncomp\ for Konus\textit{--Wind}, and \fncomp\ for \textit{Fermi}/GBM, and we use the predicted numbers for the variance term in the denominator
Since the weights of the GRBs are directly computed from the models (equations \ref{eq_nzmax0} to \ref{eq_nzmaxl}), this procedure permits the comparison of an observed quantity, the number of GRBs in each class, with the theoretical prediction of each model.
We have restricted our analysis to GRBs with \eiso\ $\ge 10^{53}$ erg, because they have weights which are not too large, indicating that we detect a significant fraction of the GRB population at these energies.
% and because all models in Table \ref{tab_LF} predict that they are in the bright end of the GRB luminosity function
Table \ref{tab_compare} gives the observed and predicted number of GRBs in each class and the mean weight of GRBs within each energy class.
The parameters of the best fit energy function are obtained with a minimization of the chisquare.\footnote{For the power law fits, we have also indicated the best fit parameters that maximize the likelihood function, showing the consistency with the chi-square analysis.}

Considering the power law fits, our analysis gives slopes that are compatible with previous works involving a power law luminosity function at high luminosity or high energy.
The value found here ($\gamma = -1.6\pm0.25$) is compatible with the values obtained by \cite{Wanderman2010} ($\gamma = -1.4^{+0.6}_{-0.3}$), \cite{Salvaterra2012} ($\gamma =-2.3^{+0.3}_{-0.8}$ for density evolution and $\gamma =-1.9^{+0.11}_{-0.14}$ for luminosity evolution), \cite{Howell2014} ($\gamma =-2.59 \pm 0.93$), or \cite{Pescalli2015} ($\gamma =-1.84 \pm 0.24$), for instance.
On the other hand, Table \ref{tab_compare} shows that the choice of a cutoff PL model leads to a shallower slope of the energy distribution ($\gamma\sim$ \textbf{-0.9 to -1.1} vs $\gamma\sim -1.6$).

\begin{figure*}
\gridline{\fig{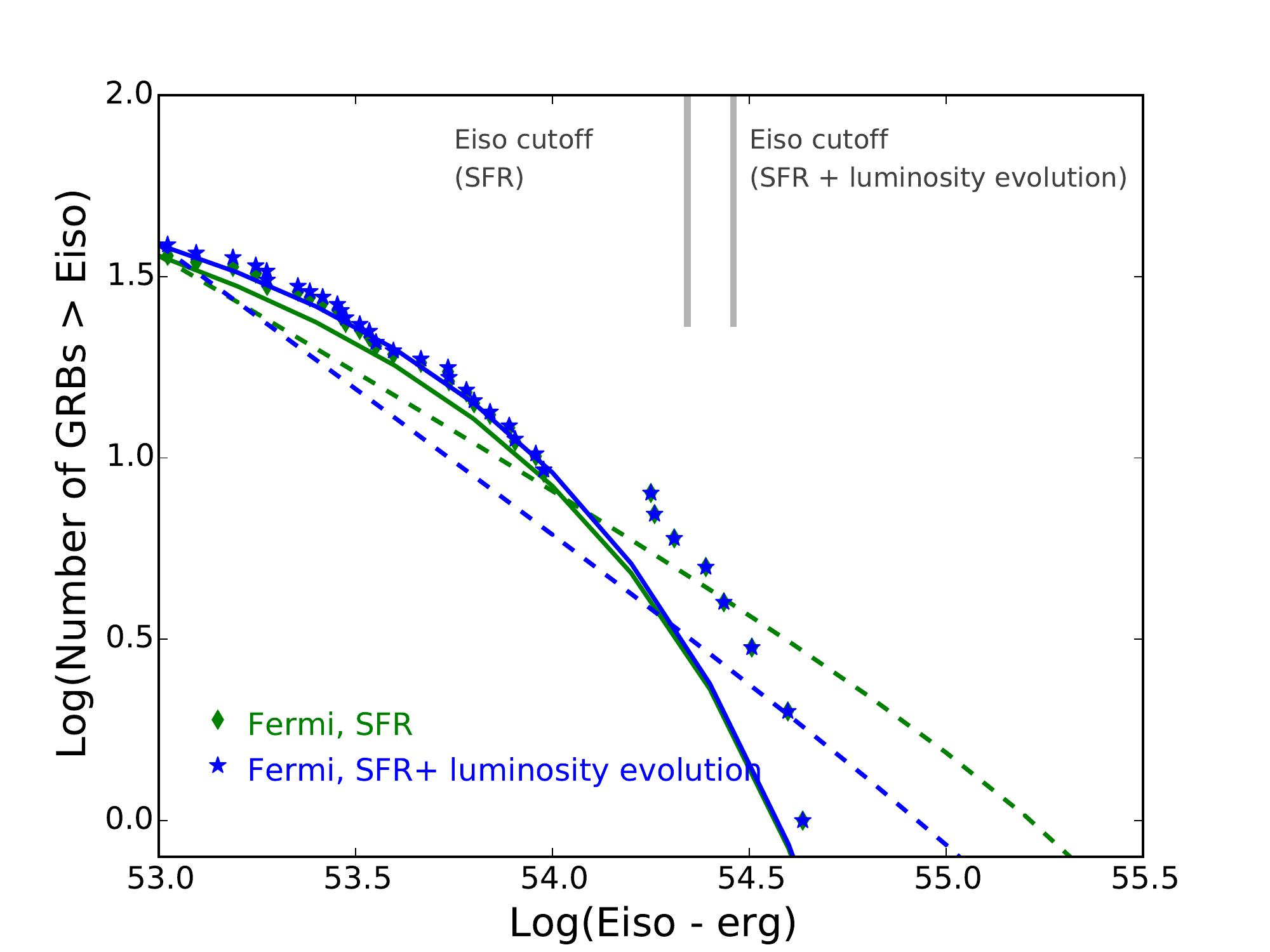}{0.45\textwidth}{(a)}  
              \fig{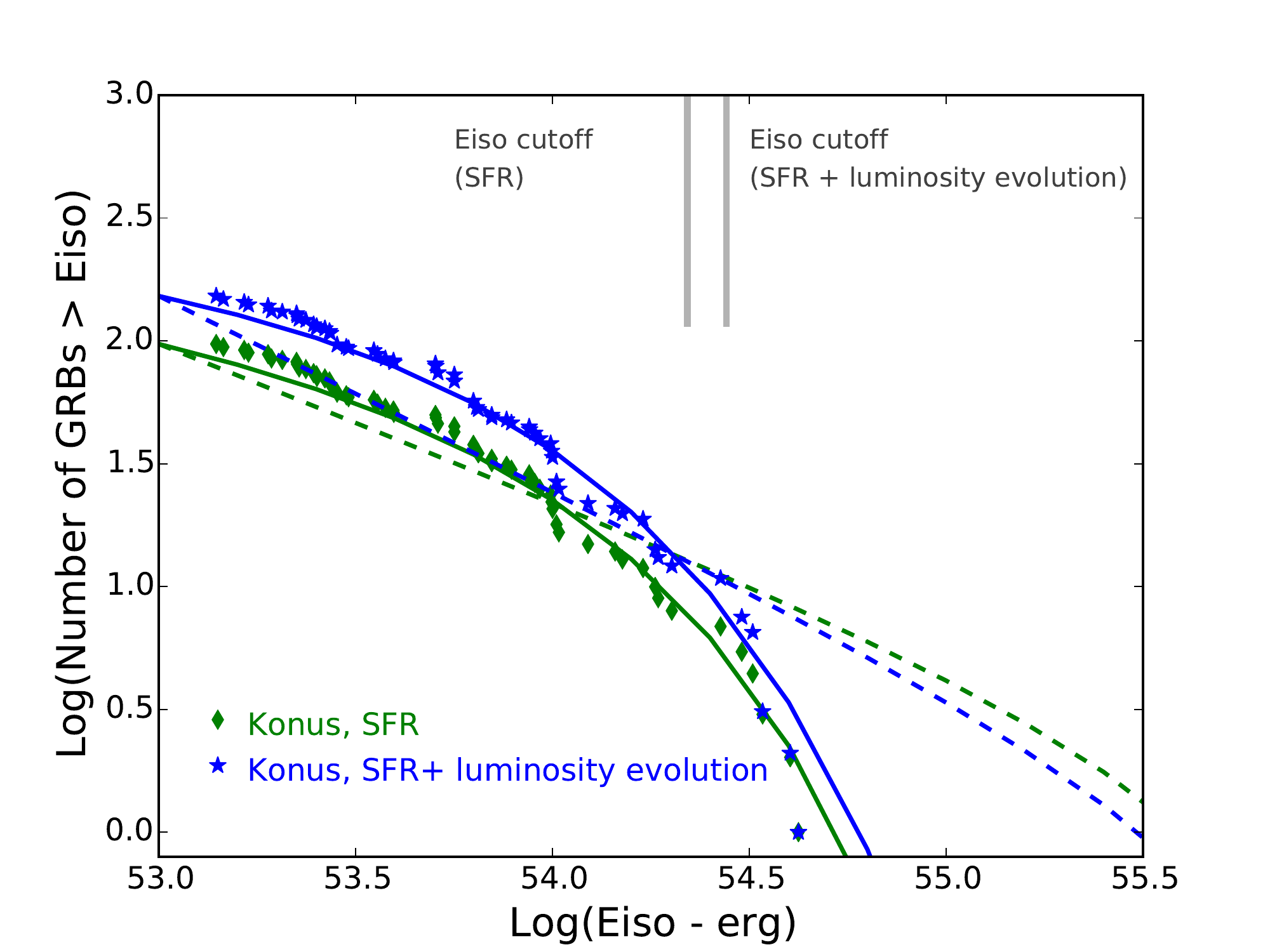}{0.45\textwidth}{(b)}}
\caption{Comparison of the best fit power law (dashed line) and of the best fit cutoff power law (solid line) \eiso\ distributions with the distributions measured by \textit{Fermi} (left) and Konus (right). We have plotted the best fit distributions for the Konus sample because they are more constrained. 
The grey vertical lines show the cutoff energy derived in Table \ref{tab_compare}, for the luminosity evolution model, the cutoff changes with the redshift, and we have plotted the cutoff energy at the median redshift of the sample (z=\fmedz\ for \textit{Fermi}/GBM and z=\kmedz\ for Konus\textit{--Wind}).
The symbols are the same as in Figure \ref{fig_EisoCDF}.}
\label{fig_models}
\end{figure*}

Our main point concerns the comparison of the simple power law energy distribution with the cutoff power law energy distribution. 
Table \ref{tab_compare} shows that the addition of the cutoff improves the fit, only slightly for the \textit{Fermi} sample, but significantly for the Konus\textit{--Wind} sample.
We attribute the larger improvement measured for Konus\textit{--Wind} to the larger number of energetic GRBs in the Konus\textit{--Wind} sample: \kncomp\ GRBs with \eiso~$\ge 10^{53}$ erg, versus \fncomp\ for the \textit{Fermi} sample, which leads to larger numbers of GRBs in the energy classes.
Since the only difference between the two models is the addition of one free parameter (the cutoff energy), the chisquare difference follows a chisquare law with one degree of freedom, allowing measuring the significance of the improvement. 
The chisquare difference $\Delta \chi \geq 10$ measured for the Konus\textit{--Wind} sample shows that the energy cutoff is required at a level larger than \textbf{99.8\%}.
We stress that the need for the GRB energy cutoff does not depend on the GRB world model, as shown by the chisquare values in Table \ref{tab_compare}.
This result is illustrated in Figure \ref{fig_models}, which compares the best fit energy distributions with the distribution of \eiso\ observed by Konus for four of the six models studied here.

In order to assess the physical reality of the cutoff, we have checked that it is not due to an instrumental effect. 
The instrumental dead time could produce a saturation of the measured flux due to the loss of photons during very bright peaks exceeding 10$^5$ cts/sec on the detector.
However, this effect cannot explain a saturation of the energy, which is an intrinsic GRB property. 
Specifically, we have checked that the most energetic GRBs in our samples are not specially bright in the observer frame (see Figure \ref{fig_EisovsX} panel f):
the six \textit{Fermi} GRBs (resp. Konus\textit{--Wind} GRBs) with \eiso~$> 2.3 \times 10^{54}$ erg have the following rank in term of their observed peak photon flux: 1-26-20-8-24-3 (resp. 47-55-50-4-11-1).
Given the count rates of these bursts, the measurements of their \eiso\ are not affected by significant dead time effects.
Since there is no mechanism that could prevent the detection of very energetic GRBs or affect strongly the measurement of \eiso , we conclude that the energy cutoff of the gamma-ray isotropic emission at $1-3 \times 10^{54}$ erg is an intrinsic property of the sources.

We also checked the energy of GRBs outside the redshift range considered here.
The most energetic GRB below z=1 is GRB~110918A at z=0.984, with \eiso $=  2.3 \times 10^{54}$ erg measured by Konus\textit{--Wind}, \citep{Frederiks2013}.
The extremely bright GRB~130427A located  at z= 0.34 stands a factor three below the limit, with \eiso $=  8 \times 10^{53}$ erg \citep[e.g.][]{Ackermann2014, Maselli2014, Perley2014, Vestrand2014}. 
The most energetic GRB above z=5 is GRB~130606A at z= 5.913, with \eiso $=  2.7 \times 10^{53}$ erg measured by Konus\textit{--Wind},  \citep{Golenetskii2013} a factor ten below the limit discussed here.
Thus, GRBs outside the redshift range [1--5] do not exceed the energy limit derived from GRBs with redshift in this range.

We finally note that ultra-long GRBs \citep[e.g.][]{Gendre2013,Levan2014}, do not exceed the energy limit discussed here despite their long duration.
GRB~111209A for instance has \eiso~=$6\times 10^{53}$ erg, four times below the cutoff energy.

\begin{deluxetable}{l l l l l l l l l l}
\tabletypesize{\footnotesize}
\tablecaption{Comparison of the observed number of GRBs with the predictions of six models.
Column 1 describes the GRB world model. 
Columns 2 to 6 give the observed and predicted number of GRBs in 5 energy classes.
Columns 7 and 8 give the parameters of the best fit energy function, based on chi-square minimization (upper row) and on maximum likelihood (lower row, only for the power law distribution). 
For models with luminosity evolution, the parameters correspond to the energy function at redshift z=0.
Column 9 indicates the agreement between the observed and predicted number of GRB based on a chi-square test.
Column 10 indicates the agreement between the observed and predicted redshift distributions, a good agreement corresponding to $\langle {\rm Nz/Nz_{max}} \rangle = 0.5$ (Section \ref{sub_GRBFR}). 
Error bars are indicated for the confidence level of 90\%.
\label{tab_compare}}
\tablehead{\colhead{Model} &\multicolumn{5}{c}{Number of GRBs in the energy range (\eiso\ in {erg})}& \colhead{Best fit} & \colhead{Cutoff} & \colhead{$\chi^2$} & \colhead{$\langle {\rm Nz/Nz_{max}} \rangle$ } \\
			& \colhead{$10^{53-53.5}$} &\colhead{$10^{53.5-54}$}& \colhead{$10^{54-54.5}$} &\colhead{$10^{54.5-55}$}& \colhead{$10^{55-56}$}	& \colhead{ PL index}	&  \colhead{$10^{54}$ erg} & (dof) & }
\startdata
\bf \textit{Fermi}/GBM, observed  &\bf  12.0 &\bf  14.0 &\bf   5.0 &\bf   3.0 &\bf   0.0 &  &  &  &  \\
\hline\hline
Model: PL, no evolution  &  15.5 &   8.9 &   4.9 &   2.6 &   2.1 & -1.55$\pm 0.20$ & N/A & 5.88 & 0.45$\pm 0.066$ \\
\textit{Fermi}/GBM mean weights  &  1.1 &  1.0 &  1.0 &  1.0 &  1.0 & -1.55$\pm 0.19$ &  & (3) &  \\ 
\hline
Model: PL \& density evol.  &  15.1 &   9.1 &   5.1 &   2.6 &   2.1 & -1.57$\pm 0.21$ & N/A & 5.41 & 0.35$\pm 0.066$ \\
\textit{Fermi}/GBM mean weights  &  1.3 &  1.1 &  1.0 &  1.0 &  1.0 & -1.57$\pm 0.18$ &  & (3) &  \\ 
\hline
Model: PL \& luminosity evol.  &  15.3 &   9.2 &   5.1 &   2.5 &   1.9 & -1.61$\pm 0.23$ & N/A & 5.11 & 0.31$\pm 0.066$ \\
\textit{Fermi}/GBM mean weights  &  1.5 &  1.1 &  1.0 &  1.0 &  1.0 & -1.59$\pm 0.18$ &  & (3) &  \\ 
\hline\hline
Model: CPL \& no evolution  &  12.8 &  11.3 &   7.5 &   2.3 &   0.1 & -1.07 &   3.3 & 1.82 & 0.45$\pm 0.066$ \\
\textit{Fermi}/GBM mean weights  &  1.1 &  1.0 &  1.0 &  1.0 &  1.0 &  &  & (2) &  \\ 
\hline
Model: CPL \& density evol.  &  12.7 &  11.4 &   7.5 &   2.3 &   0.1 & -1.13 &   3.5 & 1.81 & 0.35$\pm 0.066$ \\
\textit{Fermi}/GBM mean weights  &  1.3 &  1.1 &  1.0 &  1.0 &  1.0 &  &  & (2) &  \\ 
\hline
Model: CPL \& luminosity evol.  &  12.7 &  11.4 &   7.4 &   2.3 &   0.2 & -1.11 &   1.2 & 1.82 & 0.33$\pm 0.066$ \\
\textit{Fermi}/GBM mean weights  &  1.5 &  1.2 &  1.0 &  1.0 &  1.0 &  &  & (2) &  \\ 
\hline\hline
\bf Konus--\textit{Wind}, observed  &\bf  20.0 &\bf  23.0 &\bf  12.0 &\bf   4.0 &\bf   0.0 &  &  &  &  \\
\hline\hline
Model: PL, no evolution  &  25.7 &  15.3 &   8.7 &   5.2 &   4.1 & -1.63$\pm 0.15$ & N/A & 10.69 & 0.48$\pm 0.056$ \\
Konus--\textit{Wind} mean weights  &  2.0 &  1.6 &  1.4 &  1.1 &  1.0 & -1.57$\pm 0.12$ &  & (3) &  \\ 
\hline
Model: PL \& density evol.  &  26.0 &  15.2 &   8.5 &   5.3 &   3.9 & -1.73$\pm 0.16$ & N/A & 11.02 & 0.40$\pm 0.056$ \\
Konus--\textit{Wind} mean weights  &  3.1 &  2.3 &  1.7 &  1.2 &  1.0 & -1.61$\pm 0.10$ &  & (3) &  \\ 
\hline
Model: PL \& luminosity evol.  &  25.6 &  15.0 &   8.5 &   5.6 &   4.3 & -1.79$\pm 0.17$ & N/A & 11.76 & 0.35$\pm 0.056$ \\
Konus--\textit{Wind} mean weights  &  5.0 &  3.4 &  2.4 &  1.4 &  1.0 & -1.65$\pm 0.09$ &  & (3) &  \\ 
\hline\hline
Model: CPL \& no evolution  &  21.0 &  20.8 &  13.7 &   3.4 &   0.1 & -1.03$\pm 0.3$ &   $2.2_{-1.0}^{+4}$ & 0.63 & 0.48$\pm 0.056$ \\
Konus--\textit{Wind} mean weights  &  2.0 &  1.6 &  1.4 &  1.1 &  1.0 &  &  & (2) &  \\ 
\hline
Model: CPL \& density evol.  &  20.6 &  20.9 &  13.8 &   3.6 &   0.1 & -1.09$\pm 0.4$ &   $2.1_{-1.0}^{+4}$ & 0.59 & 0.40$\pm 0.056$ \\
Konus--\textit{Wind} mean weights  &  3.0 &  2.3 &  1.7 &  1.2 &  1.0 &  &  & (2) &  \\ 
\hline
Model: CPL \& luminosity evol.  &  20.3 &  21.8 &  13.3 &   3.5 &   0.1 & $-0.87_{-0.55}^{+0.7}$ &   $0.6_{-0.3}^{+1.2}$ & 0.39 & 0.37$\pm 0.056$ \\
Konus--\textit{Wind} mean weights  &  5.1 &  3.6 &  2.6 &  1.5 &  1.0 &  &  & (2) &  \\ 
\enddata
\tablenotetext{a}{We give no error on the best fit parameters for Fermi CPL models since they are not well constrained due to a degeneracy between \\
the slope of the power law and the cutoff energy for small numbers of GRBs.}
\end{deluxetable}

\section{Discussion}
\label{sec_discussion}

\subsection{Very energetic GRBs}
\label{sub_vegrb}
We start this section with a brief discussion of the main properties of very energetic GRBs (hereafter called "energetic GRBs" for simplicity), that we arbitrarily define as GRBs with \eiso $> 2.3 \times 10^{54}$ erg.
This cut selects the six most energetic events of each instrument.
Four energetic GRBs have been detected in common by \textit{Fermi}/GBM and Konus\textit{--Wind}: GRB~080916C, GRB~090323, GRB~120624B, and GRB~160625B.
Two have been detected only by \textit{Fermi}/GBM: GRB~090902B and GRB~140206A, and two only by Konus\textit{--Wind}: GRB~130505A and GRB~130907A. 
These energetic GRBs are bright events which are detectable out to z$\geq$5 with \textit{Fermi}/GBM, and out to distances ranging from z=2.07 (GRB~130907A) to z$\geq$5 (GRB~080916C) with Konus\textit{--Wind}.

\begin{figure*}
\gridline{\leftfig{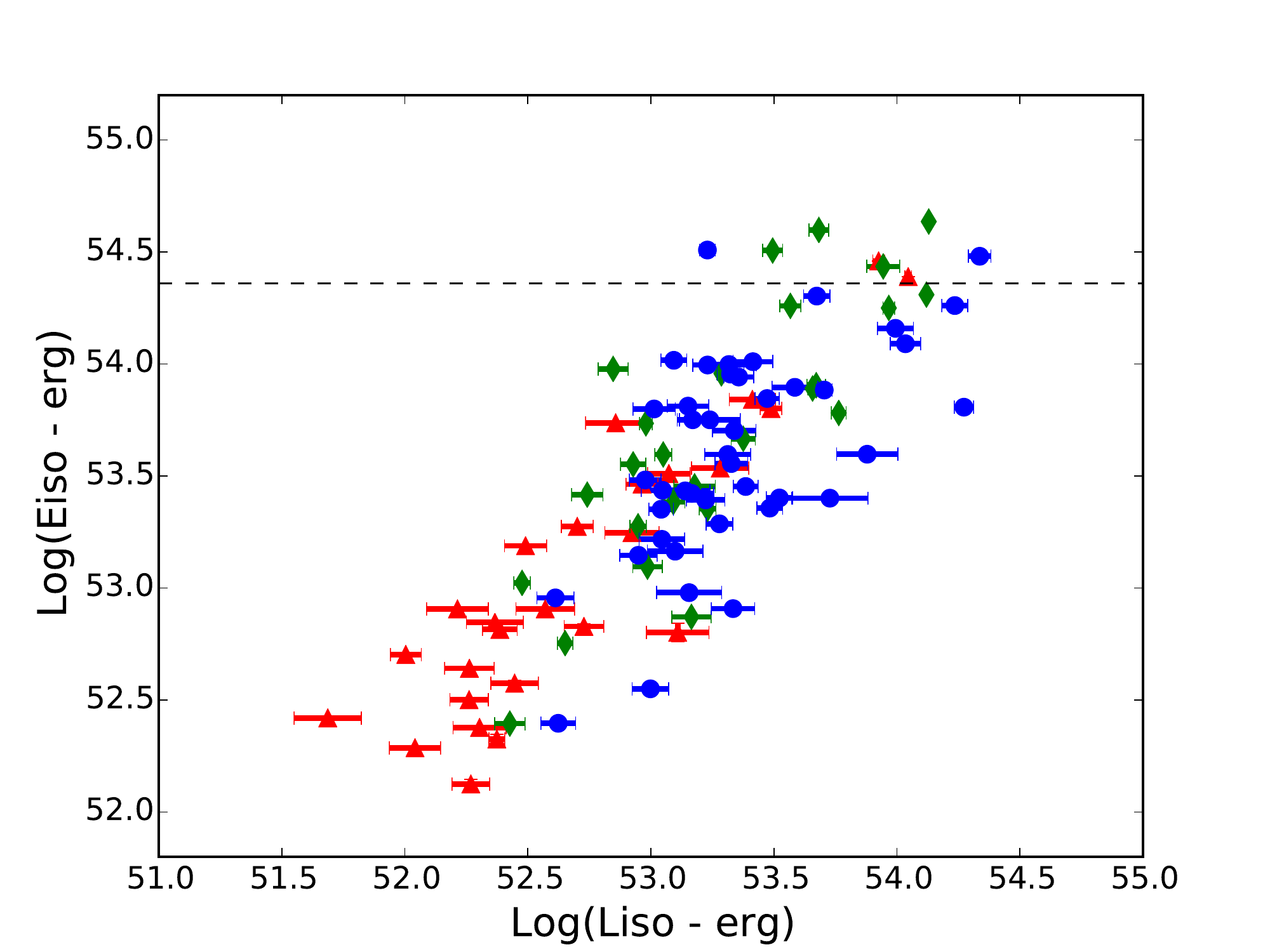}{0.45\textwidth}{(a)} % 
              \fig{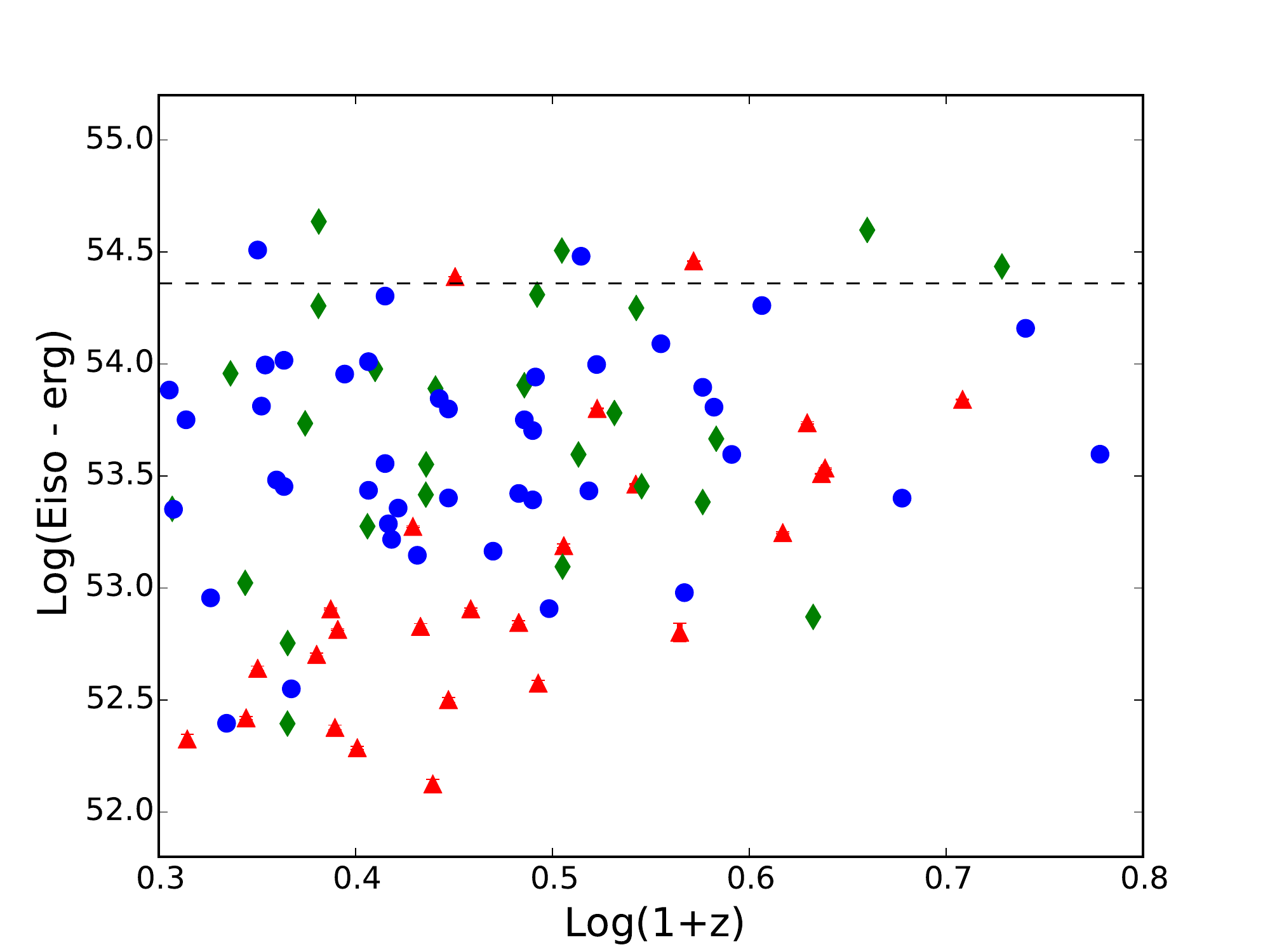}{0.45\textwidth}{(b)}} 
\gridline{\leftfig{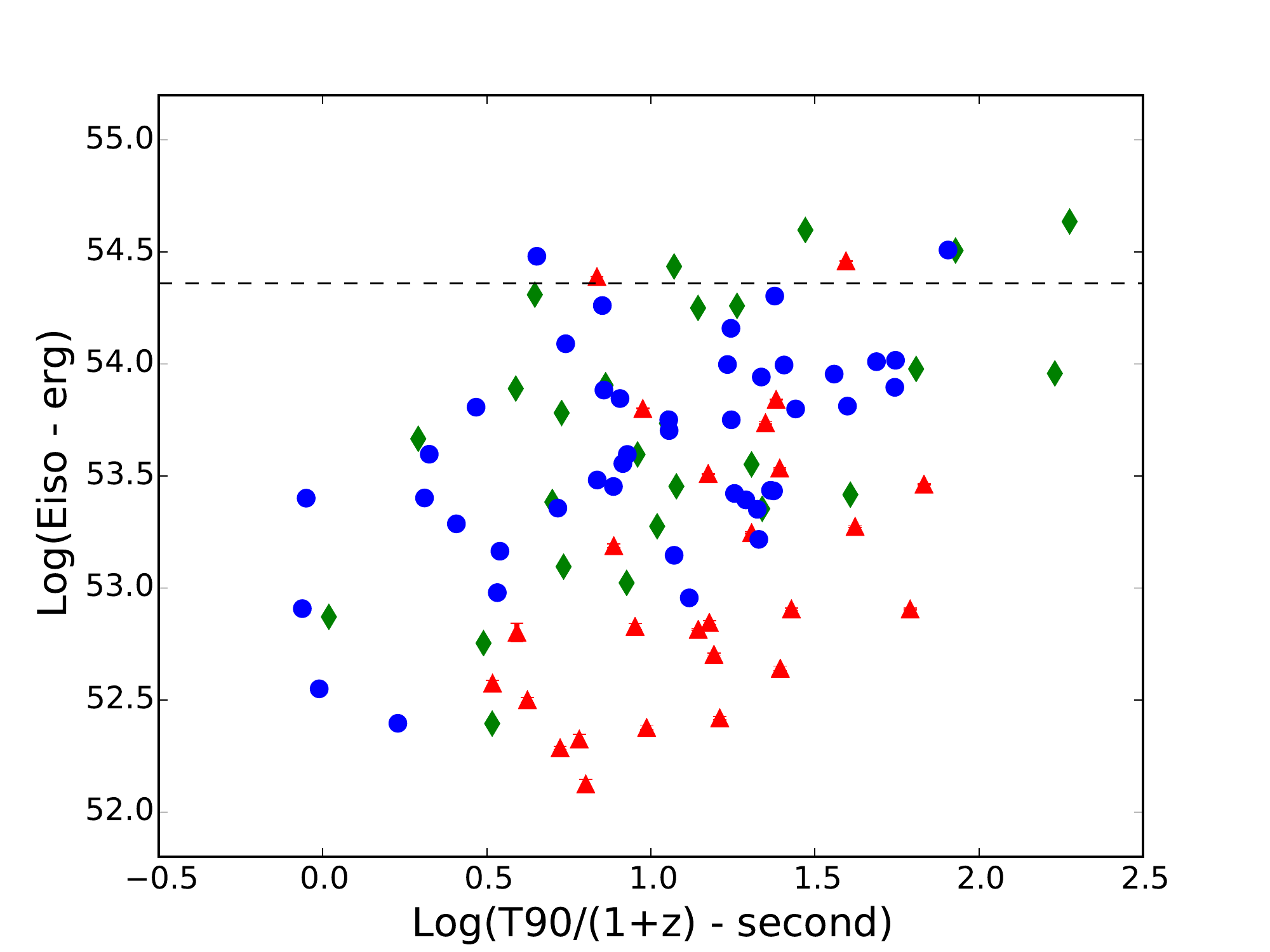}{0.45\textwidth}{(c)} % 
              \fig{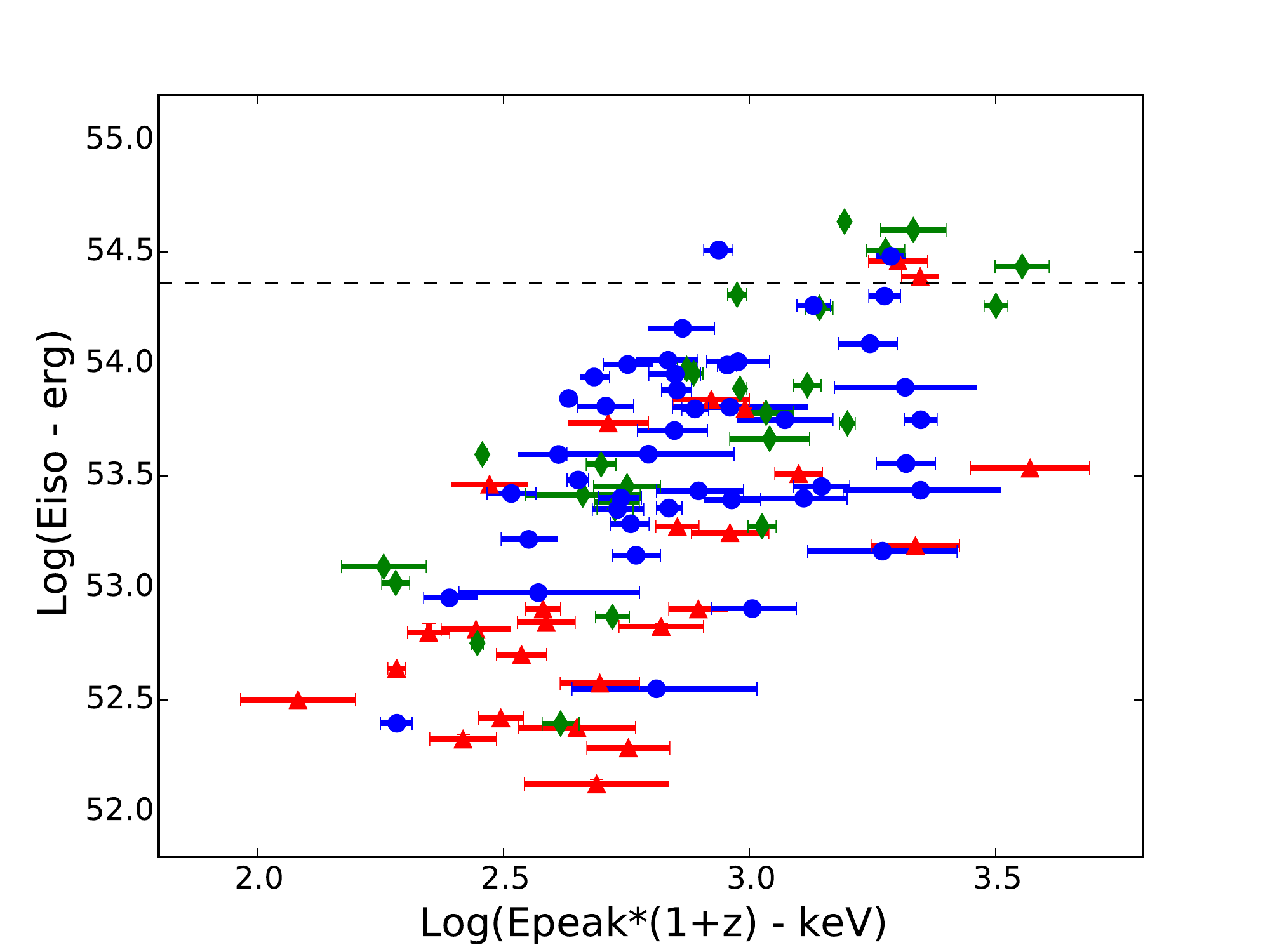}{0.45\textwidth}{(d)}} % 
\gridline{\leftfig{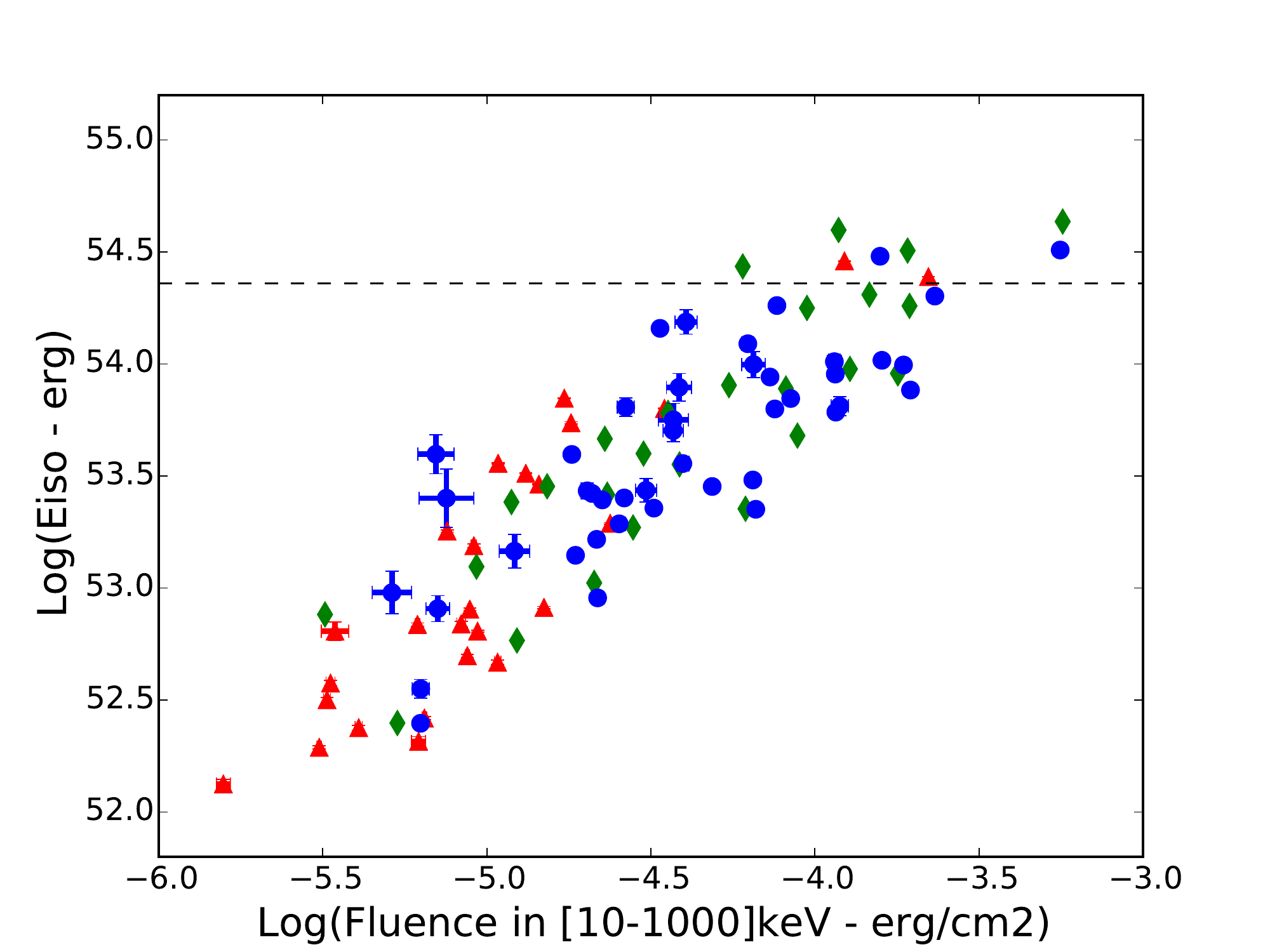}{0.45\textwidth}{(e)}% 
	      \fig{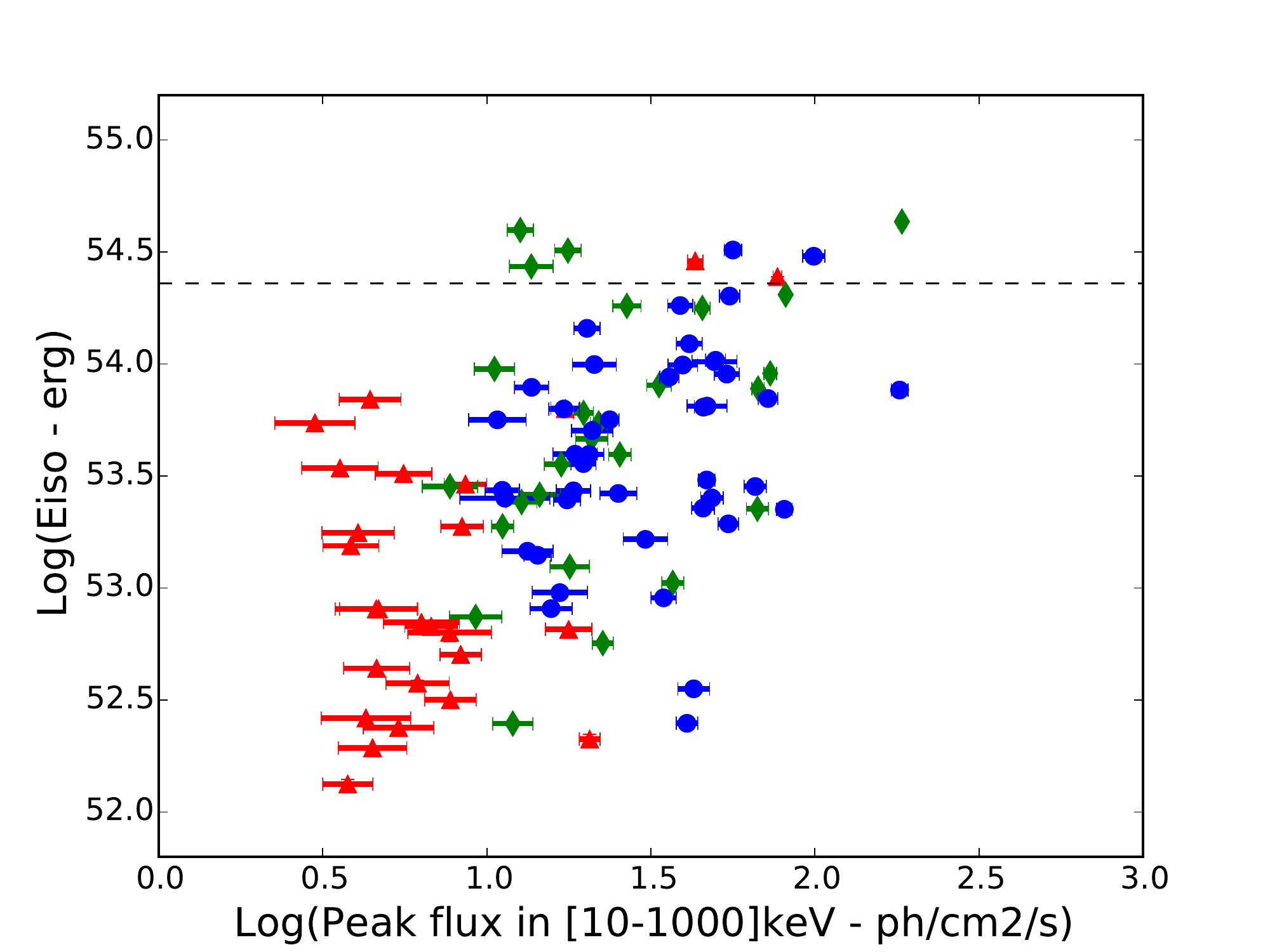}{0.45\textwidth}{(f)}}
\caption{Distribution of \eiso\ for \orngrb\ GRBs as a function of \liso\ (panel a), the redshift (panel b), \dur\ in the restframe (panel c), \epi\ in the restframe (panel d), the 10 keV - 1 MeV fluence (panel e), and the peak flux (panel f). 
We use red triangles for \textit{Fermi}-only GRBs, blue circles for Konus-only GRBs, and green diamonds for GRBs detected by \textit{Fermi} and Konus.
For those GRBs, we plot the values from the \textit{Fermi}-GBM catalog.
Energetic GRBs discussed in Section \ref{sub_vegrb} are located above the dashed line.}
\label{fig_EisovsX}
\end{figure*}

Figure \ref{fig_EisovsX} compare the properties of these eight energetic GRBs (located above the dashed line) with other GRBs in our sample.
Energetic GRBs appear longer than average, with larger fluence and larger intrinsic \epeak . 
Their intrinsic durations range from 6.4~s to 189~s, with a median of 34~s, larger than the median intrinsic duration of 11.9~s for \textit{Fermi}/GBM GRBs and of 9.1~s for Konus\textit{--Wind} GRBs. 
Their observed fluences range from $6 \times 10^{-5}$~erg~cm$^{-2}$ to $90 \times 10^{-5}$~erg~cm$^{-2}$, with a median of $21 \times 10^{-5}$~erg~cm$^{-2}$, larger than the median fluence of $1.6 \times 10^{-5}$~erg~cm$^{-2}$ for \textit{Fermi}/GBM GRBs in our sample and of $5.4 \times 10^{-5}$~erg~cm$^{-2}$ for Konus\textit{--Wind} GRBs in our sample. 
Their intrinsic \epeak\ range from 870 to 3580 keV with a median of 1850 keV, well above the median intrinsic peak energy of \textit{Fermi} GRBs (670 keV) and Konus\textit{--Wind} GRBs (730 keV).
This last feature agrees with a known property of GRBs, namely that GRBs with large \eiso\ cannot have low intrinsic \epeak\ \citep{Amati2009, Heussaff2013}.
We point out that these energetic GRBs are not specially distant sources, since their redshifts range from z=1.24 to z=4.35, with a median value z=2.2, close to the median of our sample.
Finally, we note that the six energetic GRBs detected by \textit{Fermi}/GBM have also been detected by the LAT, according to the \textit{Fermi} LAT online GRB catalog\footnote{http://fermi.gsfc.nasa.gov/ssc/observations/types/grbs/lat\_grbs/}, indicating that GeV emission is systematically detected in energetic GRBs (see also \citealt{Veres2015} about GRB~130907A). 
This means that the values of \eiso\ given in table \ref{tab_fermi} must be taken as lower limits because part of the energy is radiated above 100 MeV, in the energy range of the LAT.
However, this very high energy emission does not change our conclusion about a cutoff energy, as explained in the next section.

Overall, we have no indication that energetic GRBs constitute a special class of events, it rather seems that they represent the high energy end of the \eiso\ distribution of long GRBs (Figure \ref{fig_EisovsX}).
This is at odds with the conclusions of \cite{Cenko2011} who claim the existence of a class of hyper-energetic GRBs, containing GRB~090323, GRB~090902B and GRB~090926A included in our sample.

\subsection{Origin of the energy cutoff}
\label{sub_cutoff}
The existence of a sharp structure in the distribution of \eiso\ remains puzzling for jetted GRBs since \eiso\ depends on several parameters, like the size of the energy reservoir feeding the jet $E_j$, the radiative efficiency of the jet $\eta_j$, and the beaming factor of the jet ($f_b = {4\pi}/\Omega_j$, where $\Omega_j$ is the solid angle of the jet) according to the formula: 

\begin{equation}
\label{eq_jet}
E_{\rm iso} = {E_j \times f_b \times \eta_j}
\end{equation}

%There are several mechanisms which may explain a sharp cutoff of \eiso.
%However, there are other ways to obtain a cutoff on \eiso\ .
%One such possibility has been recently suggested by \cite{Xie2016}, who show that neutrino dominated accretion flows around black holes with a non-zero torque boundary condition provide a limit on GRB luminosity which is compatible with the most luminous and energetic GRBs.
A simple explanation to the observed cutoff in the distribution of the isotropic energy could be obtained if it could be attributed to a dominant term in equation \ref{eq_jet}:

\begin{enumerate}
\item If GRB jets have similar geometries and radiative efficiencies, or if at least the product $f_b \times \eta_j$ is similar, then the observed cutoff would mark an upper limit on $E_j$, the energy budget of the jet, i.e. an important constraint on the physics of the central engine and the relativistic ejection. 
While the estimate of $E_j$ is very uncertain, we note that for typical values of $f_b$ ($\approx 500$) and $\eta_j$ ($\approx 0.25$), the cutoff corresponds to $E_j \approx 2  \times 10^{51}$ erg, comparable to the maximum rotational energy of magnetars \citep[][and reference therein]{Bernardini2015}.
The \eiso\ cutoff observed here could thus find a natural explanation within the context of magnetar models of GRBs. 
Nevertheless, this possibility puts stringent constraints on the efficiency of jet production in magnetars, since it requires that the rotational energy is almost entirely transferred to the jet.
\item Alternatively, if the central engine of GRBs is injecting a universal energy per unit solid angle in the jet, i.e. if $E_j \times f_b$ is similar in all GRBs, then the observed cutoff would indicate a maximum radiative efficiency, leading to an important constraint on the dissipation mechanisms and radiative processes responsible for the GRB prompt emission.
\item Finally, if the true radiated energy $E_j \times \eta_j$ is similar in all GRBs, as suggested by \cite{Ghirlanda2013}, the observed cutoff would be due to a minimum beaming angle of the jet, leading again to a new constraint on the relativistic ejection mechanism.
\end{enumerate}

Unfortunately, there are no observational evidence for such simple scenarios.
While we have some indications in favor of a high radiative efficiency of energetic GRBs, with $\eta_j$ in the range [0.2-0.6] \citep{Racusin2011}, the situation is more complex with the beaming factor.
\citet{McBreen2010} find a large dispersion of beaming factors, from $f_b \leq 180$ ($\theta_j \geq 6\degr$) for GRB~080916C and GRB~090902B to $f_b \sim$1500 ($\theta_j \leq2.1\degr$) for GRB~090323.
\citet{Cenko2011}, on the other hand, find less dispersed values for the same GRBs: $f_b \sim [390-540]$ ($\theta_j = 3.9\pm0.2\degr$) for GRB~090902B to $f_b \sim [640-900]$ ($\theta_j = 2.8\degr\ _{-0.1}^{+0.4}$) for GRB~090323.
Regarding GRB~130907A, \citet{Veres2015} reach contrasted conclusions: the afterglow can be modeled with a single jet with a beaming factor $f_b \sim 45$ ($\theta_j \geq 12\degr$) or with a double jet, with the internal (more energetic) jet having a large beaming factor $f_b \sim 1600$ ($\theta_j \sim 2\degr$).
Other studies of luminous GRBs with good multi-wavelength follow-up have led to beaming angles of $f_b \sim [700-1600]$ ($\theta_j \sim 2-3\degr$) for bright GRBs \citep[e.g.][]{Frail2001,Grupe2006}, and the most luminous GRB to date, GRB~110918A have also been suggested to be highly collimated event with  $\theta_j = 1.7-3.4 \degr$, corresponding to a large beaming factor $f_b \approx 600-2200$ \citep{Frederiks2013}.
Finally, detailed studies of well observed \textit{Swift} GRBs suggest that most of them are observed off-axis, a fact that may impact these estimates \citep{Ryan2015,Zhang2015}, it is not clear however if this is also the case for the very bright GRBs discussed here.
We conclude that the data at hand are not sufficient to firmly settle the issue of the "homogeneity" of the jets of energetic GRBs.
Overall, it is surprising to realize that such bright GRBs do not benefit from follow-up observations that permit measuring their beaming factors without ambiguity.

Another source of uncertainty arises if a significant fraction of the energy escapes in another electromagnetic channel, for instance in high energy gamma-rays (several GeV).
We note that the most luminous GRBs in our sample are all detected with \textit{Fermi}/LAT, showing that their emission is not limited to the energy range of the \textit{Fermi}/GBM, and their bolometric energy would increase if we consider the flux measured with the LAT.
Suppressing the observed energy cutoff would nevertheless require that these GRBs radiate most of their energy above several tens of MeV. 
This is in contradiction with the analysis of the energetics of some of the most luminous long GRBs performed by \citet{Ackermann2013}, which shows that keV--MeV photons dominate the energetics, with 10\% or less of the total energy being radiated above 100 MeV.

In view of these various sources of error it appears quite difficult to state if the cutoff observed on \eiso\ is due to a cutoff on the jet energy $E_j$ or to some radiative or beaming effect.
Measuring $E_j$ directly through the radiocalorimetry \citep{Frail2000} of some very energetic GRBs might offer a way to settle this issue.

In conclusion, since \eiso /4$\pi$ represents the electromagnetic energy radiated per unit of solid angle, 
the observation of a cutoff \eiso\ suggests the existence of a \textit{maximum energy radiated per unit of solid angle}.
While the very energetic GRBs discussed here radiate considerable energy, they are not necessarily those with the largest energy reservoirs.  
Indeed, GRBs with larger energy reservoirs and smaller \eiso\ could exist, if they have a different radiative pattern (broader jets) or a smaller radiative efficiency.

%Despite these uncertainties, we note that the observed cutoff on \eiso\ corresponds to a maximum jet energy $E_j \sim 2 \times 10^{52}$ erg if we consider typical values of the radiative efficiency ($\eta \sim$0.25) and of the beaming factor ($f_b \sim$500). 

\subsection{A detour through the GRB formation rate}
\label{sub_GRBFR}
In this section, we complete our analysis with a comparison of the observed and predicted redshift distributions for the six GRB world models under study.
For each observed GRB we compute two numbers: N($<$z$_i$), the number of such GRBs that are closer than the redshift of the burst, and N($<$\zmax ), the number of such GRBs within the horizon \zmax .
These two numbers depend on the choice of a GRB world model.
Considering that the observed GRBs are randomly chosen among the observable GRBs, if the world model is correct the ratio N($<$z$_i$)/N($<$\zmax ) is randomly distributed in [0,1] with a mean =0.5.
The distribution of N($<$z$_i$)/N($<$\zmax ) can thus be used to test GRB world models: a mean close to 0.5 indicates a GRB world model that is acceptable, while a mean incompatible with 0.5 indicates a GRB world model which must be rejected because it predicts a redshift distribution incompatible with the observed redshift distribution.

\textbf{Column 10 of Table \ref{tab_compare} shows that similar results are obtained with \textit{Fermi}/GBM and Konus\textit{--Wind}, suggesting that models with no evolution are favored by the data. 
This is however a low significance effect, and further studies are required to assess the compatibility of specific GRB world models with the observed N($<$z$_i$)/N($<$\zmax ) distribution.
\footnote{In the earlier version of this paper, the observed contradiction between the results obtained by Fermi and Konus was due to the error on the calculation of Konus GRB horizon.}}

\subsection{Are energetic GRBs standard candles?}
\label{sub_candle}
If it is confirmed, the existence of a limit on the GRB isotropic energy would permit using energetic GRBs as standard candles visible out to large redshifts.
We briefly discuss here the expected number of such GRBs, using the statistics of \textit{Fermi} detections.
Figure \ref{fig_EisovsX} shows that the six energetic GRBs detected by \textit{Fermi} have peak fluxes larger than 10~ph~cm$^{-2}$~s$^{-1}$ in the energy range 10-1000~keV. 
We thus consider only those GRBs in the following discussion based on the Third Fermi GBM GRB Catalog \citep{NarayanaBhat2016}.
This catalog contains 247 GRBs with a peak flux larger than 10~ph~cm$^{-2}$~s$^{-1}$ in the energy range 10-1000~keV. 
40 of them have a redshift, 11 with z $< 1$ and 29 with $1 \le {\rm z} \le 5$.
 Among them 5 are energetic GRBs with \eiso $> 2.3 \times 10^{54}$ erg (we exclude GRB~160625B which is outside the six year period covered by the Third Fermi GBM GRB Catalog).
Assuming that the fraction of energetic GRBs is the same for bright GRBs with and without a redshift, we expect 5*(247/40) = 31 energetic GRBs in six years, corresponding to a rate of  $\approx 5$/yr.
These GRBs may represent an interesting tool to explore the Hubble diagram at large redshifts (z $\ge 1.5$) if the \eiso\ cutoff discussed here does not evolve with the redshift.

%However, before we can do so, it is necessary to understand the origin of this feature and its possible evolution with the redshift.

\section{Conclusion}
\label{sec_conclusion}
The main conclusion of this paper is the existence of a sharp cutoff of the \eiso\ distribution of Konus\textit{--Wind} and \textit{Fermi}/GBM GRBs around $1-3\times 10^{54}$ erg. 
Given the scarcity of such energetic GRBs, this cutoff can only be observed by instruments with a large effective sky coverage (in yr~steradian). 
This is obviously the case of Konus\textit{--Wind} launched 22 years ago, and to a lesser extent the case of \textit{Fermi} launched 8 years ago, both instruments monitoring nearly the whole sky (except 30\% occulted by the Earth for \textit{Fermi}/GBM).

We have shown that this cutoff is an intrinsic GRB property, which must be taken into account by GRB world models, which may otherwise consider a slope of the bright end of the GRB energy function which is too steep.
After discussing diverse possibilities for the origin of this feature, we conclude that it is necessary to measure the fundamental properties of the jet, like the beaming angle or the true energy budget, more accurately before we can decide if this cutoff is due to the progenitor or to the physical processes at work in the jet.

% Ajouter un mot sur le fait qu'on a une chandelle standard (qui peut Žvoluer avec z ?)

%When we take into account the sharp energy cutoff required by the data, the two samples studied here are compatible with a GRB population following the Star Formation Rate, with no additional evolution (in density or luminosity). 

\acknowledgments

DT gratefully acknowledge financial support from the OCEVU LabEx, France. 
YZ welcomes financial support from IRAP (UMR5277/CNRS/UPS).
DDF gratefully acknowledges support from RFBR grants 15-02-00532-i and 16-29-13009-ofi-m.
The authors thank the referee whose comments contributed to improve the content and clarity of the manuscript.
This article made use of the GRB table maintained by J. Greiner, available at 

\textit{http://www.mpe.mpg.de/$\sim$jcg/grbgen.html}.


%% for the paper.  Each keyword is check against the master list during
%% copy editing.  Individual instruments can be provided in parentheses,
%% after the keyword, but they are not verified.

\vspace{5mm}
\facilities{\textit{Fermi, Swift, WIND}}

\software{Python}

\end{document}